\documentclass[fleqn,usenatbib]{mnras}

\usepackage[T1]{fontenc}
\usepackage{ae,aecompl}

\usepackage{graphicx}	
\usepackage{amsmath}	
\usepackage{amssymb}	

\usepackage{newtxtext,newtxmath}

\title[SNIa yields effect]{The effects of different Type Ia SN yields on Milky Way chemical evolution}

\author[M. Palla]{
Marco Palla $^{1,2,3}$\thanks{E-mail: marco.palla@inaf.it}
\\
\\
$^{1}$Dipartimento di Fisica, Sezione di Astronomia, Universit{\'a} degli Studi di Trieste, via G. B. Tiepolo 11, I-34131, Trieste, Italy\\
$^{2}$INAF, Osservatorio Astronomico di Trieste, Via G.B. Tiepolo 11, I-34143 Trieste, Italy\\
$^{3}$INFN, Sezione di Trieste, via A. Valerio 2, I-34100, Trieste, Italy
}

\date{Accepted XXX. Received YYY; in original form ZZZ}

\pubyear{2021}

\begin{document}
\label{firstpage}
\pagerange{\pageref{firstpage}--\pageref{lastpage}}
\maketitle

\begin{abstract}
We study the effect of different Type Ia SN nucleosynthesis prescriptions on the Milky Way chemical evolution. To this aim, we run detailed one-infall and two-infall chemical evolution models, adopting a large compilation of yield sets corresponding to different white dwarf progenitors (near-Chandrasekar and sub-Chandrasekar) taken from the literature. We adopt a fixed delay time distribution function for Type Ia SNe 
, in order to avoid degeneracies in the analysis of the different nucleosynthesis channels. We also combine yields for different Type Ia SN progenitors in order to test the contribution to chemical evolution of different Type Ia SN channels. The results of the models are compared with recent LTE and NLTE observational data. We find that "classical" W7 and WDD2 models produce Fe masses and [$\alpha$/Fe] abundance patterns similar to more recent and physical near-Chandrasekar and sub-Chandrasekar models. For Fe-peak elements, we find that the results strongly depend either on the white dwarf explosion mechanism (deflagration-to-detonation, pure deflagration, double detonation) or on the initial white dwarf conditions  (central density, explosion pattern). The comparison of chemical evolution model results with observations suggests that a combination of near-Chandrasekar and sub-Chandrasekar yields is necessary to reproduce the data of V, Cr, Mn and Ni, with different fractions depending on the adopted massive stars stellar yields. This comparison also suggests that NLTE and singly ionised abundances should be definitely preferred  when dealing with most of Fe-peak elements at low metallicity. 
\end{abstract}


\defcitealias{Leung18}{L18}
\defcitealias{Leung20}{L20}
\defcitealias{Seit13}{S13}
\defcitealias{Maeda10}{M10}
\defcitealias{Travaglio04}{T04}
\defcitealias{Shen18}{S18}
\defcitealias{Iwa99}{I99}
\defcitealias{Fink14}{F14}

\section{Introduction}
\label{s:intro}
Type Ia supernovae (hereafter, SNe Ia) are one of the most important phenomena in Astrophysics. For example, they are used as "standard" candles for measuring cosmological distances (\citealt{Phillips93}) and they are beneath the discovery of the accelerating expansion of the Universe (\citealt{Riess98,Perlmutter99}). SNe Ia are also the origin of most of the iron and some of the Fe-peak elements\footnote{elements with atomic number adjacent to iron.} in galaxies (e.g. \citealt{Nomoto84,Matteucci12}).

The SN Ia events are assumed to be originated to be thermonuclear explosions by white dwarfs (WDs) in binary systems. Nonetheless, the physics governing SNe Ia is still largely debated (see \citealt{Hillebrandt00,Hillebrandt13,Maoz14,Ruiter20} for a review). In particular, the main problems reside on the progenitor system and the explosion mechanism.\\
In the last decades, the most popular proposed scenarios for the explosion mechanisms of SNe Ia have been: (i) deflagration or delayed detonation of a near-Chandrasekhar mass (near-$M_{ch}$) carbon-oxygen (CO) WD in a single degenerate system (e.g. \citealt{Whelan73}), (ii) near-Chandrasekar mass explosion in a double degenerate system (e.g. \citealt{Iben84}), (iii) double detonations of sub-Chandrasekar mass (sub-$M_{ch}$) WD in a single or double degenerate system (e.g. \citealt{Nomoto82b,Iben91}), (iv) violent merger in double degenerate sub-$M_{ch}$ WDs (e.g. \citealt{Pakmor12}), (v) weak deflagration of a near-$M_{ch}$ WD in a single degenerate system giving rise to low mass WD remnant. The latter scenario could correspond to a Type Iax supernova (SN Iax, see e.g., \citealt{Kromer15}).\\
Nearly all classes of explosions are supported by lightcurve and abundance observations of several individual SN explosions and remnants (see \citealt{Kirby19,DeLosReyes20}), suggesting that SNe Ia explode through multiple channels (e.g. \citealt{Mannucci06}).\\

Chemical evolution can be regarded as a valuable tool to constrain the dominant SN Ia channels by looking at elemental abundance ratios. In fact, while the nucleosynthetic distinction between single and double degenerate channels may not be large enough to be constrained with current data, the mass of the WD (i.e. near-$M_{ch}$ or sub-$M_{ch}$) has a large effect on the production of certain elements (\citealt{Koba20}). Important roles in element production are also played by other WD features, such as the WD central density (e.g. \citealt{Leung18}) or the explosion pattern (e.g. \citealt{Seit13,Leung20}).
In the last decade, a great number of simulations devoted to study the nucleosynthesis of the different SN Ia channels have become available in the literature, with a broad exploration of the parameter space.

So far,  most of the papers regarding chemical evolution still adopt old 1D simulations, in particular the W7 model (e.g. \citealt{Nomoto84,Iwa99}) or less frequently the WDD2 model (\citealt{Iwa99}).  The yields obtained from these simulations reasonably reproduce the Galactic chemical evolution in the solar neighbourhood for most of the elements. However, these yields show some problems.  As for example, Ni is overproduced in W7, while Cr is overproduced in WDD2 (\citealt{Leung18}).\\
Up to now, only few studies of chemical evolution have been devoted to asses the impact of different SN Ia progenitor yields on the evolution of the abundance ratios (e.g. \citealt{Cescutti17,Koba20}).
However, most of these studies look at single elements (e.g. \citealt{Eitner20}) or external galaxies (mainly Milky Way satellites and other dwarfs, e.g. \citealt{Koba15,Cescutti17}). An exception is represented by \citet{Koba20}, where O, Mn, Cr and Ni abundances in the Galaxy are compared with two SN Ia models adopting sub-$M_{ch}$ and near-$M_{ch}$ channels.\\

The aim of this paper is to compare the yield outcome of a large compilation (more than 20) of SN Ia models for different nucleosynthesis channels (near-$M_{ch}$, sub-$M_{ch}$, SNe Iax), by adopting detailed models for Milky Way (MW) chemical evolution. In this way, we explore how different progenitors and different parameters affect chemical abundance ratios. To do so, we adopt a fixed delay time distribution function (DTD) and in particular that pertaining to the single degenerate model.  The possible effects of different DTDs are beyond the scope of this paper, since our focus is on the SN Ia nucleosynthetic yields. By the way, the DTD of double degenerate systems, as well as observationally inferred ones, are similar to the one adopted here (see \citealt{Matteucci09}). \\
We adopt either a one-infall (\citealt{Grisoni18}) and a revised two-infall model (\citealt{Palla20b}) applied to the solar vicinity. In the case of the two-infall model, we allow to have contribution from multiple progenitor channels, in order to test the role of the different SN Ia subclasses in MW chemical evolution.

The paper is organised as follows. In Section \ref{s:yields}, we list and describe the SN Ia models considered in this paper. In Section \ref{s:models}, we present the chemical evolution models adopted in this work. In Section \ref{s:results}, we first show the chemical evolution results for the different yield sets and later we present the comparison between model predictions and observations, also discussing the role of the different SN Ia channels. Finally, in Section \ref{s:conclusions}, we draw our conclusions.

\section{SN Ia yields}
\label{s:yields}
In this section we introduce the different SN Ia yield sets adopted in this work. The reader is referred to the source papers for a more thorough discussion of the adopted input physics.\\

We sample different explosion mechanisms and progenitor masses. In particular, we look at delayed detonation/deflagration-to-detonation transition (hereafter DDT) and pure deflagrations (PTD) models for near-$M_{ch}$ WD, while for sub-$M_{ch}$ WD we adopt double-detonation (DD) models.\\
A compilation of the models adopted in this work can be found in Table \ref{t:models}, where we list the references, the explosion mechanisms and main features of the models. In the following subsections, we will briefly describe the main features of the models adopted.

\begin{table*}
    \centering
    \caption{SN Ia models.  Horizontal lines divide the standard models adopted in chemical evolution (W7 and WDD2), deflagration-to-detonation transition models (DDT), pure deflagration models (PTD) and double detonation models (DD).  For each yield set, the main parameters of the benchmark model are listed.}
    \begin{tabular}{c | c  c  c}
        \hline
        \hline
        Model  & Authors & Explosion, WD mass & Main properties\\
        \hline
         W7 \citetalias{Leung18} & \citet{Leung18} & PTD, near-$M_{ch}$ & 1D, parameters fine-tuned to match observation, different Z available\\[0.1cm]
         WDD2 \citetalias{Leung18} & \citet{Leung18} & DDT, near-$M_{ch}$ & 1D, parameters fine-tuned to match observation \\
         
        \hline
        DDT \citetalias{Seit13} & \citet{Seit13} & DDT, near-$M_{ch}$ & 3D, different number of ignition sites, WD $\rho_c$ and Z available \\
        & & & bench: $M_{WD}$=$1.40$ | $\rho_c$=$2.9\cdot10^9$ | C/O$\simeq$1 | ignit. sites=100 | Z=0.01, 0.1, 0.5, 1  \\[0.1cm]
        DDT \citetalias{Leung18} & \citet{Leung18} & DDT, near-$M_{ch}$ & 2D, centred ignition, different WD $\rho_c$ and Z available\\
        & & & bench: $M_{WD}$=$1.38$ | $\rho_c$=$3\cdot10^9$ | C/O=1 | Z=0, 0.1, 0.5, 1, 2, 5 \\
        \hline
        PTD \citetalias{Fink14} & \citet{Fink14} & PTD, near-$M_{ch}$ & 3D, different number of ignition sites and WD $\rho_c$ available\\
        & & & bench: $M_{WD}$=$1.40$ | $\rho_c$=$2.9\cdot 10^9$ | C/O$\simeq$1 | ignit. sites=100 | Z=1  \\[0.1cm]
        PTD \citetalias{Leung18} & \citet{Leung18} & PTD, near-$M_{ch}$ & 2D, centred ignition, different WD $\rho_c$ available \\
        & & & bench: $M_{WD}$=$1.38$ | $\rho_c$=$3\cdot 10^9$  | C/O=1 | Z= 1  \\
        \hline
        DD \citetalias{Shen18} & \citet{Shen18} & DD, sub-$M_{ch}$ & 1D, bare CO WD, different WD masses, Z and C/O available \\
        & & & bench: $M_{WD}$=$1.00$ | $M_{He}$=$0.00$ | C/O=1  | Z=0, 0.25, 0.5, 1 \\[0.1cm]
        DD \citetalias{Leung20} & \citet{Leung20} & DD, sub-$M_{ch}$ & 2D, different WD masses, He shell masses, detonation patterns and Z available \\
        & & & bench: $M_{WD}$=$1.00$ | $M_{He}$=$0.05$ | C/O=1 | det. pattern=spheric | Z=0, 0.1, 0.5, 1, 2, 5 \\
         \hline
         \hline
    \end{tabular}
    \label{t:models}
    {\bf Notes.} $M_{WD}$, $M_{He}$ are expressed in M$_\odot$. $\rho_c$ is expressed in gr cm$^{-3}$. Z is expressed in Z$_\odot$.
\end{table*}

\subsection{W7 and WDD2 models}
\label{ss:W7_WDD2}
As aforementioned in Section \ref{s:intro}, most of galactic chemical evolution studies still adopt W7 or WDD2 models.  These models were the first able to describe with success the SNe Ia contribution to chemical evolution (e.g. \citealt{Matteucci85,Matteucci86,Koba06}) as well as the features of Type Ia SN light curves and spectra (e.g. \citealt{Hoeflich96}).\\
The models differentiate for the presence or absence of the detonation transition: the W7 model (e.g. \citealt{Nomoto84,Thielemann86,Iwa99}) adopts a PTD scheme, while WDD2 (\citealt{Iwa99}) is a DDT model.

Despite of their success, these models suffer substantial physical limitations. In fact, some model parameters (i.e. propagation flame speed, density at the detonation transition) are fine-tuned to reproduce the observables in both W7 and WDD2. Moreover, the 1D modelling represents an important limitation, since the deflagration burning front is highly textured and non-sphericity is thus actually essential (e.g. \citealt{Niemeyer96}).\\
For these reasons more realistic multi-dimensional (multi-D) models have to be preferred. In fact, in multi-D models different outcomes can be imposed only by different initial conditions (e.g. WD structure and first ignition place).

Anyway, we will show the results of W7 and WDD2 models for comparison with the other yields used in this work. In particular, we will show the updated W7 and WDD2 models presented in \citet{Leung18}, with a refined nuclear reaction network relative to those of \citet{Iwa99}.

\subsection{Multi-D DDT models}
\label{ss:DDT}
The deflagration-to-detonation transition (DDT) mechanism was introduced by \citet{Khokhlov91} to overcome some shortcomings of pure detonation (no production of intermediate-mass elements) and pure deflagration models (e.g. overproduction of neutron rich nuclear species, see \citealt{Seit17} for more information). 
In these models, an initial subsonic deflagration front burns and expands the WD.  After some time delay (and some suitable conditions) the deflagration turns into a supersonic detonation that burns the remaining fuel.

For what concerns models with the DDT mechanism, we consider two recent studies that sample a variety of different WD initial conditions. In particular, we test the yields from \citet{Seit13} and \citet{Leung18}.

\subsubsection{Seitenzahl et al. (2013)}
\citet{Seit13} completed the first study in literature in which detailed nucleosynthesis for DDT in 3D was performed.\\
The authors computed the results for twelve 3D models with different number of initial deflagration sites (from 1 to 1600) at solar metallicity and with a WD central density of $\sim 3\times10^9$ gr cm$^{-3}$. For the case of 100 ignition points (N100) the set includes the results for a range of metallicity (0.01$\le Z/Z_\odot \le$1) and central density of the CO WD ($1\times10^9\le \rho_c/$gr cm$^{-3}$ $\le5.5\times10^9$).

We will consider as benchmark models the N100 ones with $\rho_c\sim 3\times10^9$ gr cm$^{-3}$ at different metallicities (see Table \ref{t:models}). Throughout the paper, we will also explore the effects on the yields of different WD central densities. We will not consider instead the results for the models with a different number of ignition sites: in fact, in most of them the nucleosynthesis is not in line with that of typical SNe Ia (e.g. $M_{^{56}Ni}\gtrsim1$M$_\odot$ or Si/Fe$\sim$1).

\subsubsection{Leung \& Nomoto (2018)}
In \citet{Leung18} more than twenty 2D DDT models with central point ignition were computed, exploring broadly the parameter space. In particular, the effects of different WD central density ($0.5\times10^9\le \rho_c/$gr cm$^{-3}$ $\le5\times10^9$), metallicity (0$\le Z/Z_\odot \le$5), flame shape and turbulent flame formula were tested.\\
It has to be noted that in this paper the WD progenitor masses are extended down to the range of 1.30-1.35 M$_\odot$, which may be considered as sub-Chandrasekar masses (\citealt{Leung18}). The C ignition in such progenitors would be possible by shock compression due to surface He detonation (e.g. \citealt{Arnett96}), that however may not produce a C detonation (as in DD models, see Section \ref{ss:subMch}) due to the relatively large WD mass.

In our work, we will consider the models with the full metallicity range (0$\le Z/Z_\odot \le$5) covered, i.e. the low density ($\rho_c=10^9$ gr cm$^{-3}$,  $M_{WD}=1.33$M$_\odot$), benchmark ($\rho_c=3\times10^9$ gr cm$^{-3}$,  $M_{WD}=1.38$M$_\odot$) and high density ($\rho_c=5\times10^9$ gr cm$^{-3}$,  $M_{WD}=1.39$M$_\odot$) models. In particular, we will focus on the outcomes of the benchmark model, showing the results of the other models where the differences in the abundances are important.

\subsection{Multi-D PTD models}
\label{ss:PTD}
The pure deflagration (PTD) models had been for long time considered as the favoured models for SN Ia explosions (\citealt{Seit17}). In the PTD scenario, a subsonic flame (deflagration) allows the WD to respond to the nuclear energy release with expansion to lower densities. In this way, the burning can also produce intermediate-mass elements (IME) and not only iron-group elements (IGE)\footnote{For IME we mean elements with atomic number between Na and Ca. For IGE we mean elements with atomic number near to Fe.} as for a purely detonating WD.\\
Nowadays, pure deflagration models have been instead suggested as a possible model for peculiar subluminous SNe Ia, i.e. SNe Iax (e.g. \citealt{Kromer15,Leung18,Kirby19}). For this reason, the simulated properties (light curve, spectrum, nucleosynthesis) of these models may not be applicable to "normal" SNe Ia, as done previously in literature (i.e. with the W7 model).

As for DDT models, also for PTD models we consider two studies from the literature. In particular, we test the yields from \citet{Fink14} and \citet{Leung18}.

\subsubsection{Fink et al. (2014)}
In this paper, the authors computed PTD simulations based on the code adopted by \citet{Seit13}.  \citet{Fink14} also simulated between 1 and 1600 sites of ignition for the WD, but the models do not experience a transition to detonation (as for \citealt{Seit13}). For the N100def model (100 ignition sites), variations in WD central density ($1\times10^9\le \rho_c/$gr cm$^{-3}$ $\le5.5\times10^9$) are also explored. The metallicity is solar for all the fourteen models presented in the paper.

We decide to take the N100def as the benchmark model in our paper. In this way, we can directly see the effects of a missing detonation relative to \citet{Seit13} yields.\\
We note that \citet{Fink14} stated that models with lower number of ignition sites are more suitable to explain typical SN Iax lightcurves (e.g. SN 2005hk). For this reason we will also consider the outcomes of the N10def (10 ignition sites) model. However, despite of the lower Fe production, the variations relative to the N100def model are limited ($\lesssim 0.1$ dex) for most of the abundance ratios.

\subsubsection{Leung \& Nomoto (2018)}
Together with the DDT models presented in Section \ref{ss:DDT}, in this paper are also presented four 2D models in which the detonation transition trigger is "switched-off". These models can be seen as approximations of a failed DDT caused by some external effects (see \citealt{Leung18}).\\
The models examine how the WD central density (and hence its mass) influences the nucleosynthetis. This is done in the same range of densities explored for the \citet{Leung18} DDT models ($0.5\times10^9\le \rho_c/$gr cm$^{-3}$ $\le5\times10^9$). The metallicity of the models is solar.

In our work, we will mainly show the results for the model 300-1-C3-1P ($\rho_c=3\times10^9$gr cm$^{-3}$, see Table \ref{t:models}). However, we will also look at the impact of WD mass variation on the chemical evolution where the differences between the models are important.

We point out that in \citet{Leung20b} PTD models for CO WDs are tested specifically on SNe Iax. In these models the mass trapped by the WD remnant, that may be originated by Type Iax subclass, was also computed.  However, the resulting abundance patterns are generally very similar to those of the PTD models of \citet{Leung18} that we consider in this work.

\subsection{sub-$M_{ch}$ DD models}
\label{ss:subMch}
In SNe Ia originating from sub-$M_{ch}$ WD, C detonation is the responsible of the observed $^{56}$Ni mass. In most of sub-$M_{ch}$ models, the C detonation is triggered by a surface detonation of He (which can be in small or significant amount, depending on the model scenario): for this reason, we refer to sub-$M_{ch}$ models as double-detonation (DD) models. \\
In such a scenario, the lower central density ($\lesssim$10$^8$gr cm$^{-3}$) of the progenitor leads the C detonating WD to produce an amount of $^{56}$Ni consistent with what observed in typical SN Ia spectra ($\sim$0.5-0.7 M$_\odot$, see Figure 1 of \citealt{Seit17}).\\
  
In order to explore the impact of sub-$M_{ch}$ SN Ia progenitors on the chemical evolution, we test the yield sets from \citet{Shen18} and \citet{Leung20}.\\
The two papers reflect two different frameworks for sub-$M_{ch}$ WD progenitor. \citet{Shen18} adopted the dynamically driven double-detonation (DD) process,  which requires a small He shell mass to trigger the C-detonation in the centre: this can be approximated by a bare CO WD detonation. In \citet{Leung20} instead, the WD encloses a low but non negligible $^4$He mass ($\ge 0.05$ M$_\odot$). This latter scenario allows for both single and double degenerate progenitors. At the contrary, the dinamically driven DD allows for double degenerate progenitors only.

\subsubsection{Shen et al. (2018)}
In \citet{Shen18} 1D, spherically symmetric, central ignited detonations of bare CO WDs were simulated. Different models were run for different metallicities (0$\le Z/Z_\odot \le$2) and different WD masses (0.8$\le M_{WD}/$M$_\odot \le$1.1). Variations in the C/O ratio are also explored. \\
It has to be noted that the adoption of 1D models in the case of detonation only is not an issue as for models experiencing a deflagration. In fact,  in the case of a pure detonation we are not dealing with a heat transfer due to diffusion or convection,  which are responsible of a highly textured burning front.

In our work, we will mainly consider the metallicity dependent, 1M$_\odot$, C/O=1 model. However, we will also see the effects on the nuclesoynthesis of different WD masses. We will not discuss instead the impact of different C/O ratios, since we found it negligible (always $<< 0.1$ dex on the resulting abundance ratios).

\subsubsection{Leung \& Nomoto (2020)}
Using the same code as for 2D DDT models, \citet{Leung20} exploded the simulated WDs using double detonation (DD). In this work, \citet{Leung20} studied the effects of different metallicities (0$\le Z/Z_\odot \le$5), WD masses ($0.9\le M_{WD}/$M$_\odot \le 1.2$), as well as He shell masses ($0.05\le M_{He}/$M$_\odot \le 0.2$) and shape of the initial He detonation configuration (bubble, ring, spherical).

In our work, we will focus on the three metallicity dependent models with different He detonation configurations. In particular, we will mainly show the outcomes for the model with spherical detonation (see Table \ref{t:models}),  extending to the other models where the results are not similar.\\
For \citet{Leung20} models, we will not focus on mass variation effects, since these are similar to the ones produced by \citet{Shen18} models.

\section{Milky Way chemical evolution models}
\label{s:models}

In this Section, we present the chemical evolution models adopted in this work. To follow the chemical evolution of the solar neighbourhood we adopt two different models, in order to stress the differences between SN Ia yields or better reproduce the observational trend for solar vicinity stars.

The models are as follows:
\begin{enumerate}
    \item one-infall model (e.g. \citealt{Matteucci89,Grisoni18}). It assumes that the solar vicinity forms by means of a single gas infall episode, with a timescale of $\tau\simeq7$ Gyr. This timescale is fixed by reproducing the G-dwarf metallicity distribution in the solar vicinity (\citealt{Matteucci12}).
    \item revised two-infall model (\citealt{Palla20b}). It assumes that the MW disc forms by means of two distinct infall episodes: the first one forms the halo-thick disc, whereas the second (delayed and slower) infall gives rise to the thin disc. The infall timescales are $\tau_1\simeq1$ Gyr and $\tau_2\simeq7$ Gyr, respectively. Relative to "classical" two-infall models (\citealt{Chiappini97,Romano10}), the second infall is delayed by 3.25 Gyr instead of 1 Gyr. The assumption of a much more delayed second infall allows us to reproduce large survey data (\citealt{Palla20b,Spitoni21}), as well as asteroseismic stellar ages (\citealt{Spitoni19,Spitoni20}) in the solar neighbourhood.
\end{enumerate}

\subsection{Basic assumptions}
\label{ss:model_equations}

The basic equations that describe the evolution of a given chemical element $i$ are:
\begin{equation}
\Dot{G}_i (t) = -\psi(t) X_i(t) + R_i(t) + \Dot{G}_{i,inf}(t),
    \label{e:chemical_evo}
\end{equation}
where $G_i (t) = X_i (t) G(t)$ is the fraction of gas mass in the form of an element $i$ and $G(t)$ is the fractional mass of gas. The quantity $X_i(t)$ represents the abundance fraction in mass of a given element $i$, with the summation over all elements in the gas mixture being equal to unity.

The first term on the right hand side of Equation \eqref{e:chemical_evo} corresponds to the the rate at which an element $i$ is removed from the ISM due to star formation. We parametrise the SFR according to the Schmidt-Kennicutt law (\citealt{Kennicutt98}):
\begin{equation}
\psi(t) = \nu \Sigma_{gas}(t)^k,
    \label{e:SFR}
\end{equation}
where $\Sigma_{gas}$ is the surface gas density, $k=1.5$  is  the law index and $\nu$ is the star formation  efficiency. 

The second term in Equation \eqref{e:chemical_evo} (see \citealt{Palla20} for the complete expression) takes into account the nucleosynthesis from low-intermediate mass stars (LIMS, $m < 8$M$_\odot$), core collapse (CC) SNe (Type II and Ib/c, $m > 8$M$_\odot$) and SNe Ia. \\
The stellar yields from normal stars are taken from \citet{Karakas10} (LIMS) and \citet{Koba06} (CC-SNe).\\
As in many previous papers (e.g. \citealt{Romano10,Grisoni18,Spitoni19}) for SNe Ia we adopt the single degenerate (SD) delay-time-distribution (DTD) function from \citet{Matteucci01}:
\begin{equation}
    R_{Ia}(t)=A_{Ia}\int_{M_{B,inf}(t)}^{M_{B,sup}(t)} \phi(M_B)\, f\bigg(\frac{M_2(t)}{M_B}\bigg)\, \frac{dM_B}{M_B},
    \label{e:DTD}
\end{equation}
where $M_B$ is  the  total  mass  of  the  binary system giving rise to the SN Ia, $M_2$ is the mass of the secondary star, $M_{B,inf}=max(2 M_2, 3$ M$_\odot)$ and $M_{B,sup}=8$ M$_\odot+M_2$ the minimum  and maximum masses for the binary systems contributing at the time $t$. $f(\frac{M_2}{M_B})$ is the distribution function of the mass fraction of the secondary (see \citealt{Matteucci01} for details). $A_{Ia}$ is the parameter representing the fraction of binary systems able to produce a SN Ia and its value is set to reproduce the observed rate of SNe Ia in the Galaxy.
As can be seen in Equation \eqref{e:DTD}, in this formalism the clock for the explosion is given by the lifetime of the secondary star.\\
In \citet{Matteucci09}, it has been shown that this DTD is very similar to that related to the double degenerate model of \citet{Greggio05}.  The same happens for observationally inferred $t^{-\sim1}$ DTDs (e.g. \citealt{Totani08,Maoz17}). In fact, \citet{Totani08} showed that a $t^{-1}$ DTD is very similar to that of \citet{Greggio05}.  For this reason, even though the \citet{Matteucci01} DTD is only representative of the SD scenario, it can be considered an acceptable compromise to describe the delayed pollution from the entire SN Ia population. Moreover, we remember that our focus in this work is on SN Ia nucleosynthesis rather than on the different DTDs.\\

Concerning the initial mass function (IMF), the adopted IMF is the \citet{Kroupa93} one, derived for the solar vicinity.

The  last  term  in  Equation \eqref{e:chemical_evo}  is  the gas infall rate. For the two-infall model, the gas accretion is computed in this way:
\begin{equation}
    \Dot{G}_{i,inf}(t)=A\,X_{i,inf}\,e^{-\frac{t}{\tau_1}} + \theta(t-t_{max}) B\,X_{i,inf}\, e^{-\frac{t-t_{max}}{\tau_2}},
    \label{e:infall}    
\end{equation}
where $G_{i,inf}(t)$ is the infalling material in the form of element $i$ and $X_{i,inf}$ is the composition of the infalling gas, which is assumed to be primordial. $\tau_1$ and $\tau_2$ are the infall timescales for the first and the second infall episodes, while $t_{max}$ indicates the time of maximum infall, which is also the delay between the first and the second infall. The coefficients $A$ and $B$ are obtained by reproducing the present-day surface mass density of the the thick and thin discs in the solar neighbourhood. We also remind the reader that the $\theta$ in the Equation above is the Heavyside step function.\\
For the one-infall models, the formula gets simpler:
\begin{equation}
    \Dot{G}_{i,inf}(t)=B\,X_{i,inf}\,e^{-\frac{t}{\tau}},
    \label{e:infall_2}
\end{equation}
where the different quantities has the usual meaning.

Both the models do not include galactic winds. Galactic fountains more likely occur in galactic discs and it was found (e.g. \citealt{Melioli09,Spitoni09}) that they do not modify significantly the chemical evolution of the disc as a whole.

\section{Results}
\label{s:results}
In this Section we discuss, element by element, the behaviour of several abundance ratios as functions of metallicity ([X/Fe]\footnote{[X/Y]=$\log$(X/Y)-$\log$(X$_\odot$/Y$_\odot$), where X, Y are abundances in the ISM and X$_\odot$, Y$_\odot$ are solar abundances.} vs. [Fe/H]) for the different yield sets adopted in this work. 

We show the results for $\alpha$-element and Fe-peak element abundances. For these latter, which are the main focus of this work, we show the results for the elements in which SNe Ia production is important, i.e. V, Cr, Mn and Ni.\\
In addition, we also see which combinations of different SN Ia progenitor yields could explain the observed abundance ratios in the solar neighbourhood.

\subsection{Individual yield sets}
\label{ss:single_yields}
Here we show the contribution to chemical evolution of the different yield sets presented in Table \ref{t:models}.\\
The Section is divided into two parts: in the first one, we test the effects of different SN Ia yield sets on [$\alpha$/Fe] ratios, in which we can directly see the SNe Ia contribution to Fe production ($\alpha$-elements are underproduced by this SN class). In the second part instead, we concentrate on the Fe-peak abundances for which the SN Ia production is also relevant.

\begin{figure}
    \centering
    \includegraphics[width=1.\columnwidth]{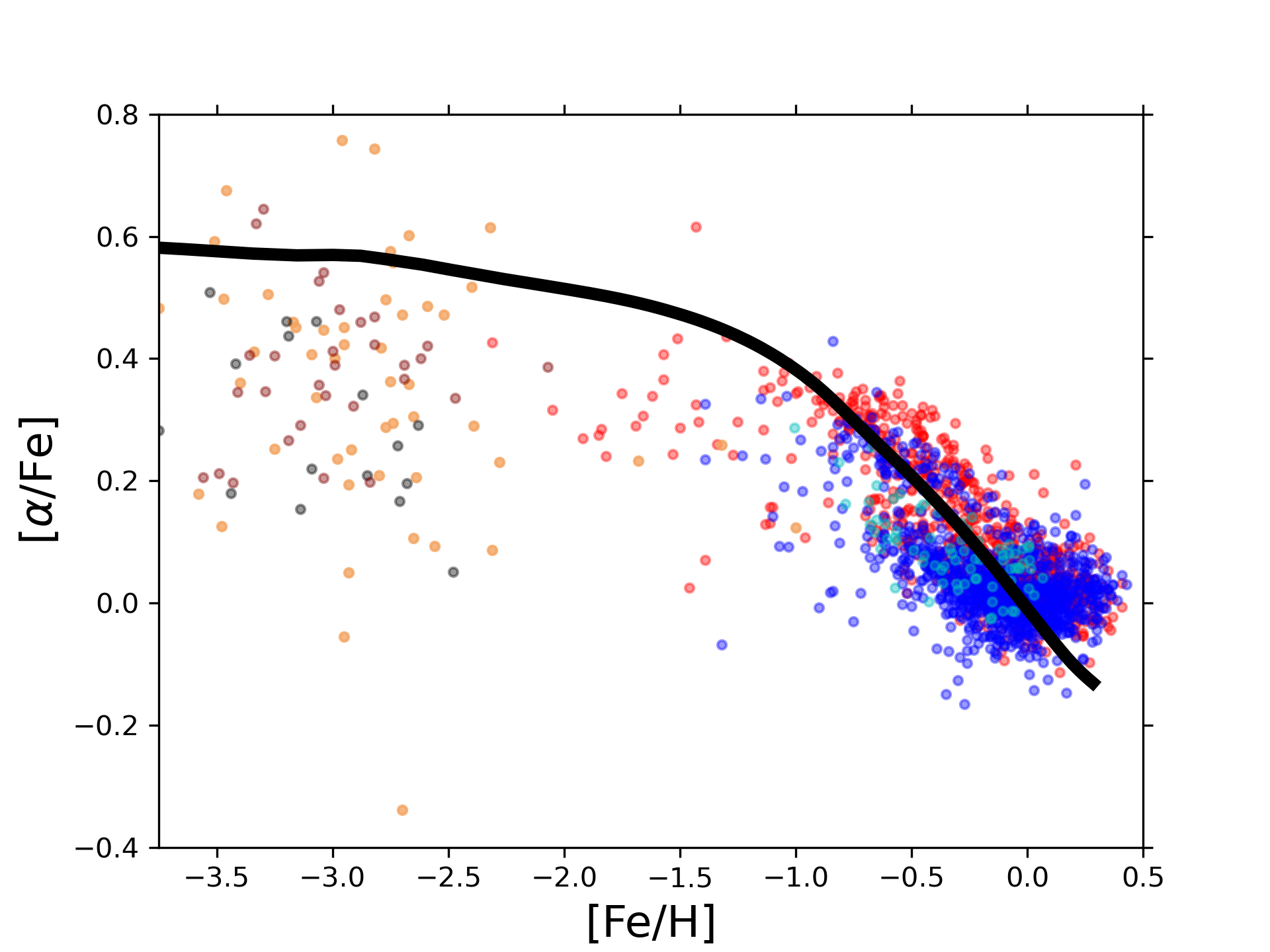}
    \caption{[$\alpha$/Fe] vs. [Fe/H] predicted by our one-infall chemical evolution model adopting standard W7 \citet{Iwa99} SN Ia yields. Data are from \citet{Chen00} (cyan points),  \citet{Cayrel04} (maroon points), \citet{Lai08} (grey points), \citet{Adibekyan12} (blue points), \citet{Yong13} (orange points) and \citet{Bensby14} (red points).}
    \label{f:alphaFe_calibr}
\end{figure}

For this first analysis, we adopt the one-infall model presented in Section \ref{s:models}.  This model allows to highlight and explain better the effects produced by the different SN Ia yields on the abundance patterns, which is the aim of this Section.  The finding of the best models to fit the observational data is reserved to the next Section, where the more physical two-infall model is adopted.\\
However,  the one-infall model is able to explain the observed [$\alpha$/Fe] vs. [Fe/H] behaviour in the solar neighbourhood both for metal poor and metal rich stars. This is shown in Figure \ref{f:alphaFe_calibr}, where a model adopting the standard W7 \citet{Iwa99} yield set is compared to the data from \citet{Cayrel04,Lai08,Yong13} (metal poor stars) and \citet{Chen00,Adibekyan12,Bensby14} (moderate to metal rich stars).\\
We note a large spread in low metallicity data, falling both below and above the model track. This can be explained by two facts. Low [$\alpha$/Fe] data are probably accreted stars from MW satellites that merged with the Galaxy during its first phase of formation (\citealt{Helmi20}). The spread around the model track is instead due to the inhmogeneous mixing that affect MW evolution at high redshift (e.g. \citealt{Cescutti08}).
Anyway, this plot shows clearly the "time-delay model" (\citealt{Matteucci03,Matteucci12}).  In fact, the Fe pollution from SNe Ia is evident for [Fe/H]$\gtrsim-1$ dex, where the contribution of SNe Ia to chemical enrichment starts to be comparable to that of CC-SNe.

\subsubsection{$\alpha$-elements}
We choose to show the results for magnesium and silicon, since they are two elements where SN Ia contribution is different.\\
Mg is negligibly produced by SN Ia, since its production comes from hydrostatic carbon buning and explosive neon burning in massive stars (e.g. \citealt{WW95}): in this way, the [Mg/Fe] ratio can be used as a direct indicator of Fe pollution from SNe Ia. On the other hand, Si receives a non negligible contribution from SNe Ia.

\subsubsection*{Magnesium}
\begin{figure}
    \centering
    \includegraphics[width=1.\columnwidth]{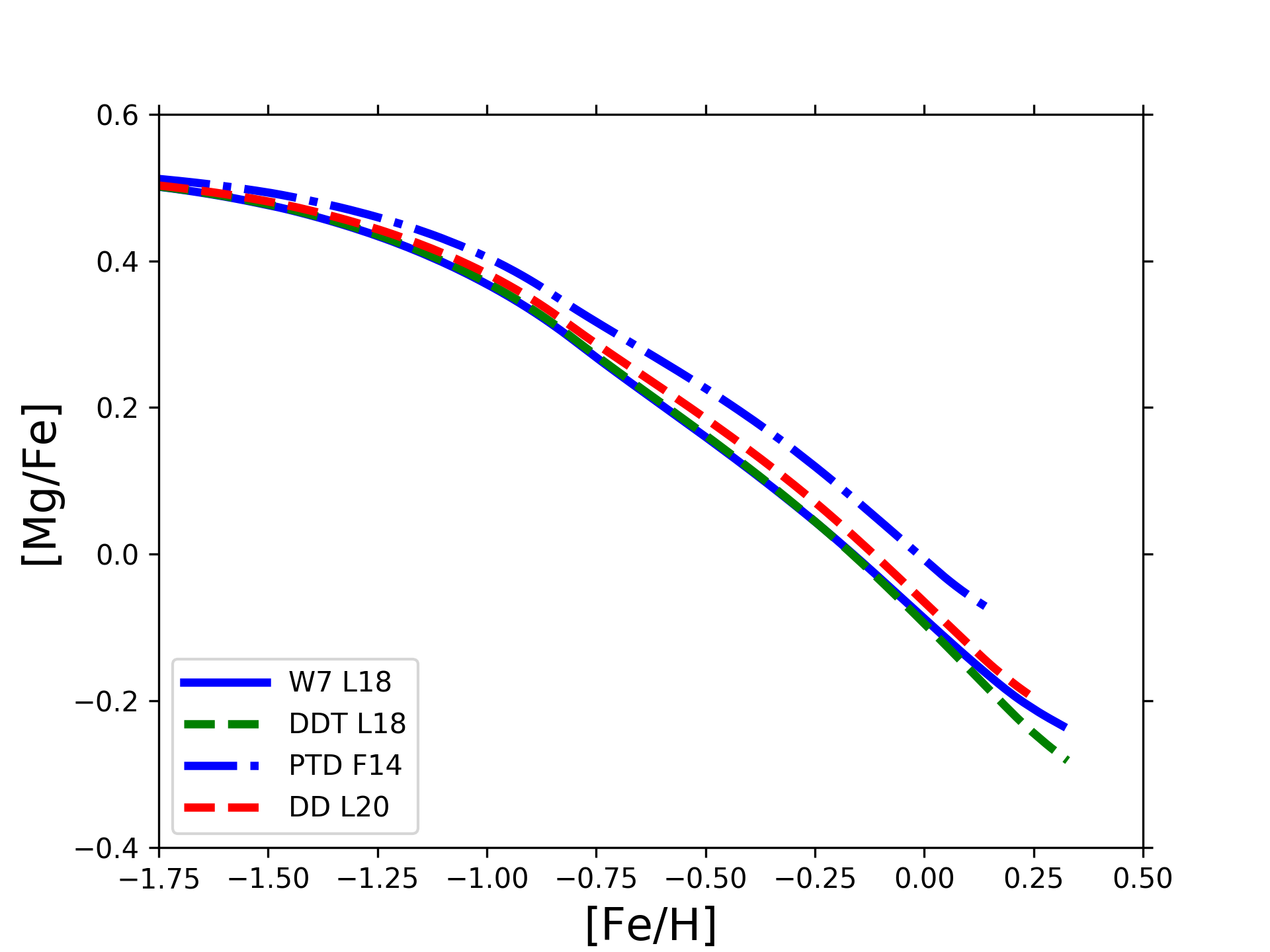}
    \caption{[Mg/Fe] vs. [Fe/H] ratios predicted by our one-infall chemical evolution model adopting different SN Ia yield sets. Results are shown for W7 \citetalias{Leung18} (blue solid line), DDT \citetalias{Leung18} (green dashed), PTD \citetalias{Fink14} (blue dash-dotted) and DD \citetalias{Leung20} (red dashed) yields.}
    \label{f:MgFe}
\end{figure}

In Figure \ref{f:MgFe} we show the model results for one DDT model (\citealt{Leung18}), one PTD model (\citealt{Fink14}) and one sub-$M_{ch}$ DD model (\citealt{Leung20}). In addition, we also plot the abundance pattern obtained using the revised W7 \citetalias{Leung18} model.  The choice of plotting just few of the model results is made to avoid overcrowding of lines in the Figure. We also plot the results only for [Fe/H]>-1.75 dex since for lower metallicities the effect of SNe Ia on abundance ratios is negligible. 
The lines color, style and thickness are organised to simply identify the explosion mechanism, the geometry and the other parameters affecting SN Ia yields.  In particular,  we plot the PTD models in blue,  the DDT models in green and the DD models red,  whereas the geometry is treated by plotting solid lines for 1D models, dashed lines for 2D models and dash-dotted lines for 3D ones.  In addition, we consider the effect of different central densities (for DDT and PTD models) and explosion patterns (for DD models) by varying the thickness of the lines and giving a shaded effect to the lines not belonging to benchmark sets.
The same organisation is adopted for the subsequent Figures. 

In Figure \ref{f:MgFe} we see that the different yield sets provide negligible variations in the abundance ratios, except for the model adopting yields with pure deflagration explosion (blue dashed line in the Figure). For this latter, we see slightly higher [Mg/Fe] at a certain [Fe/H]: this is due to lower Fe production in PTD models, in which large amounts of C and O remain unburnt (\citealt{Leung18}). We highlight that a similar behaviour is seen for the other PTD yields tested in this work (\citealt{Fink14,Leung18}).\\
The differences in iron production can be also seen by looking at Table \ref{t:Fe_abundances}, where we list Fe and other Fe-peak elements solar abundances predicted by the revised two-infall model (\citealt{Palla20b}, remember Section \ref{s:models}) adopting the different yield sets tested in this work. We choose to compare the two-infall model to solar photospheric abundances since, as explained in Section \ref{s:models}, this model reproduces in detail the features observed in the solar vicinity. For Fe, we see that benchmark DDT and sub-$M_{ch}$ DD models agree well within 0.1 dex relative to the observed solar abundance (taken from \citealt{Asplund09}), whereas PTD yields give lower predicted abundances. However, we have to remind that pure deflagration models are usually considered to be not representative of the whole SN Ia population (e.g. \citealt{Kromer15}).\\
Still regarding the predicted Fe abundance, it must be pointed out that \citet{Shen18} low-mass sub-$M_{ch}$ models (0.8M$_\odot$) produce extremely low amount of iron (<0.1M$_\odot$) to be considered as valuable SN Ia progenitor candidates. At the same time, high-mass WD models (1.1 M$_\odot$) show too large $^{56}$Ni feedback ($>0.8$M$_\odot$). Similar results for low and high mass sub-$M_{ch}$ WD are found for \citet{Leung20} models (see also Figure 9 of \citealt{Koba20}). For this reason, in the rest of the paper we will not consider mass variations for the sub-Chandrasekar models.

\begin{table*}
    \centering
    \caption{Predicted $\log$(X/H)+12 solar abundances for Fe-peak elements by our two-infall model using the different SN Ia yield sets adopted in this paper. The assumed benchmark models are written in bold. Model abundances are taken at $t=9.25$ Gyr in order to take the time at which the protosolar cloud was formed. The predictions are compared to observed photospheric solar abundances by \citet{Asplund09}. Horizontal lines divide the standard models adopted in chemical evolution (W7 and WDD2), DDT, PTD and DD models.}
    \begin{tabular}{c | c  c  c  c  c  }
        \hline
        \hline
        Observation  & $\log$(Fe/H)+12 & $\log$(V/H)+12 & $\log$(Cr/H)+12 & $\log$(Mn/H)+12 & $\log$(Ni/H)+12  \\
        \hline
        \citet{Asplund09} & 7.50$\pm$0.04 & 3.93$\pm$0.08 & 5.64$\pm$0.04 & 5.43$\pm$0.04 & 6.22$\pm$0.04 \\
        \hline
        Model & $\log$(Fe/H)+12 & $\log$(V/H)+12 & $\log$(Cr/H)+12 & $\log$(Mn/H)+12 & $\log$(Ni/H)+12\\
        \hline
        W7 \citetalias{Leung18} &  7.50 &  3.64 & 5.58 & 5.52 & 6.36\\
        WDD2 \citetalias{Leung18} & 7.48 & 3.77 & 5.79 & 5.35  & 6.25\\
        \hline
        {\bf DDT \citetalias{Seit13}} & {\bf 7.48} & {\bf 3.64} & {\bf 5.64} & {\bf 5.46}  & {\bf 6.37} \\
        DDT \citetalias{Seit13} (low $\rho_c$) & 7.43 & 3.69 & 5.62 & 5.55 & 6.24\\
        DDT \citetalias{Seit13} (high $\rho_c$) & 7.52 & 3.79 & 5.74 & 5.57 & 6.45\\
        {\bf DDT \citetalias{Leung18}} & {\bf 7.50} & {\bf 3.64} & {\bf 5.61} & {\bf 5.40}  & {\bf 6.38}\\
        DDT \citetalias{Leung18} (low $\rho_c$) & 7.50 & 3.59 & 5.54 & 5.18 & 6.28\\
        DDT \citetalias{Leung18} (high $\rho_c$) & 7.52 & 3.96 & 5.85 & 5.52 & 6.42\\
        \hline
        {\bf PTD \citetalias{Fink14}} & {\bf 7.37} & {\bf 3.53} & {\bf 5.46} & {\bf 5.45}  & {\bf 6.36} \\
        PTD \citetalias{Fink14} (low $\rho_c$) & 7.33 & 3.48 & 5.41 & 5.26  & 6.21\\
        PTD \citetalias{Fink14} (high $\rho_c$) & 7.38 & 3.68 & 5.62 & 5.49  & 6.38\\
        PTD \citetalias{Fink14} (10 ignition sites) & 7.26 & 3.44 & 5.36 & 5.18 & 6.19\\
        {\bf PTD \citetalias{Leung18}} & {\bf 7.37} & {\bf 3.52} & {\bf 5.48} & {\bf 5.37}  & {\bf 6.34}\\
        PTD \citetalias{Leung18} (low $\rho_c$) & 7.30 & 3.43 &  5.32 & 5.10 & 6.20\\
        PTD \citetalias{Leung18} (high $\rho_c$) & 7.44 & 3.92 & 5.79 & 5.50 & 6.39\\
        \hline
        {\bf DD \citetalias{Leung20} (spherical detonation)} & {\bf 7.44} & {\bf 3.61} &  {\bf 5.43} & {\bf 4.85} & {\bf 6.13}\\ 
        DD \citetalias{Leung20} (bubble detonation) & 7.45 & 4.23 & 5.78 &  4.95 & 6.18\\
        DD \citetalias{Leung20} (ring detonation) & 7.48 & 3.97 & 5.41 & 4.78 & 6.17\\
        {\bf DD \citetalias{Shen18} (1 M$_\odot$)} & {\bf 7.42} & {\bf 3.55} & {\bf 5.61} & {\bf 5.04} & {\bf 6.11}\\        
        DD \citetalias{Shen18} (0.8 M$_\odot$) & 7.13 & 3.48 & 5.36 & 4.80  & 5.95 \\
        DD \citetalias{Shen18} (1.1 M$_\odot$) & 7.51 & 3.51 & 5.54 & 4.96 & 6.22\\
        DD \citetalias{Shen18} (1 M$_\odot$, C/O=0.3) & 7.41 & 3.56 &  5.63 & 5.06  & 6.08\\
         \hline
         \hline
    \end{tabular}\\
    \label{t:Fe_abundances}
\end{table*}

\subsubsection*{Silicon}
\begin{figure}
    \centering
    \includegraphics[width=1.\columnwidth]{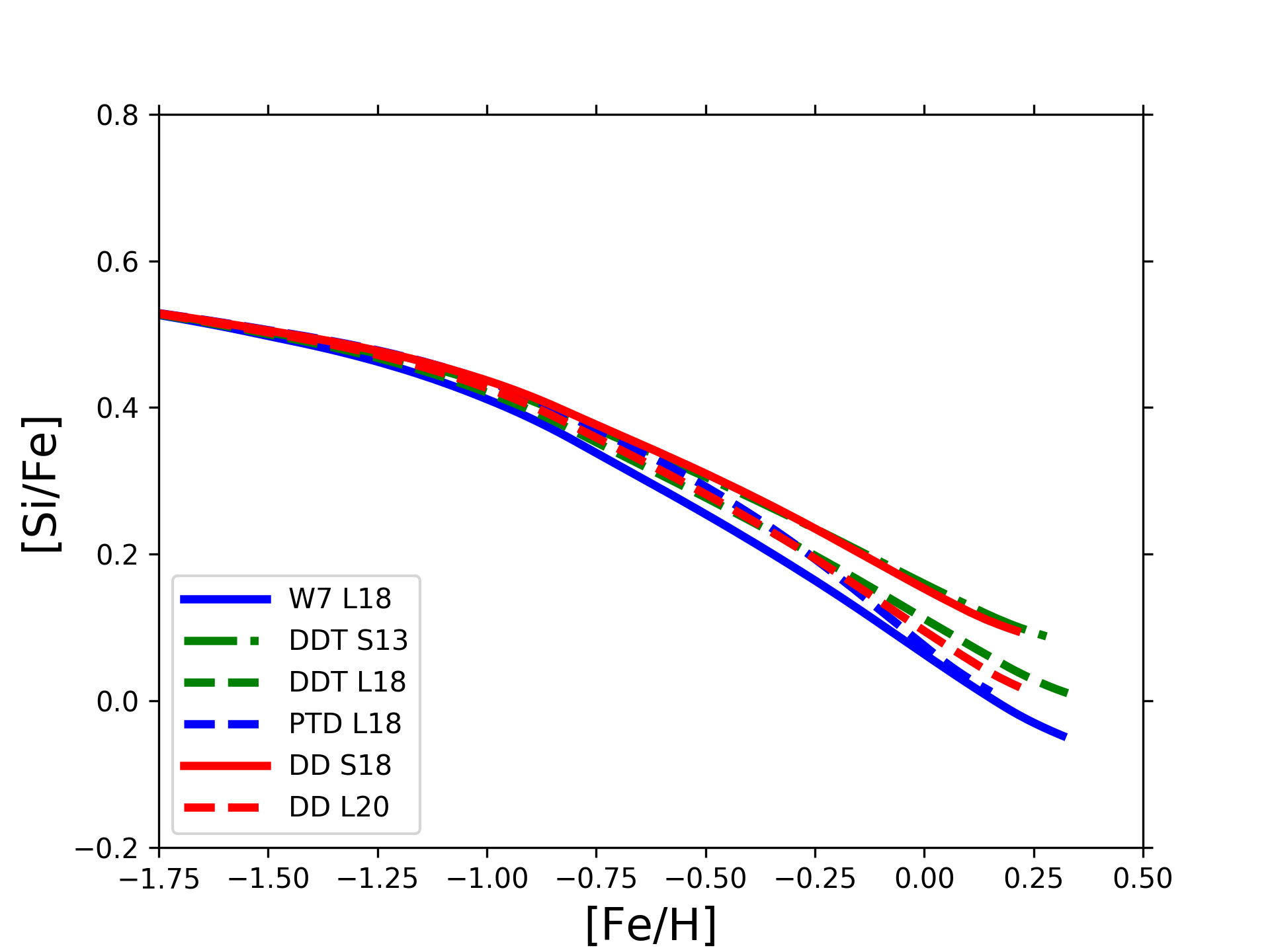}
    \caption{Same of Figure \ref{f:MgFe}, but for [Si/Fe]. Results are shown for W7 \citetalias{Leung18} (blue solid line), DDT \citetalias{Seit13} (green dash-dotted), DDT \citetalias{Leung18} (green dashed), PTD \citetalias{Leung18} (blue dashed), DD \citetalias{Shen18} (red solid) and DD \citetalias{Leung20} (red dashed) yields.}
    \label{f:SiFe}
\end{figure}

In Figure \ref{f:SiFe} we can see the effects of the different yield sets on the [Si/Fe] vs. [Fe/H] plot.\\
At variance with Mg, we see that in this case there is not a distinction on the results based on the explosion mechanism. In fact, sub-$M_{ch}$ yields of \citet{Shen18} provide a larger [Si/Fe] yield relative to those from \citet{Leung20} (for all the different He detonation trigger). The reason of a higher Si yield can be found in a more incomplete burning that leads to larger production of IMEs and light Fe-peak elements (\citealt{Leung20,Koba20}). Also DDT and PTD models show visible differences in the resulting [Si/Fe] vs. [Fe/H] relations. However, the described variations in the final abundance ratios are limited to the order of $\sim$0.1 dex.\\ 
For what concerns PTD models, we note that [Si/Fe] abundance tracks are similar, where not lower, to those of the models with other explosion mechanisms. This means that the [Si/Fe] yields for PTD models are similar to than of DDT or sub-$M_{ch}$ DD models: this means that the lack of Fe is compensated by a smaller Si production.
A similar general picture can be found for other IMEs (e.g. S, Ca). In fact, most of IME isotopes are produced during the detonation phase (see Figure 7 of \citealt{Leung18}) .

\subsubsection{Fe-peak elements}
SNe Ia have a central role in Fe-peak elements pollution. In fact, for some of these elements SNe Ia represent the main site of production.

\begin{figure}
    \centering
    \includegraphics[width=1.\columnwidth]{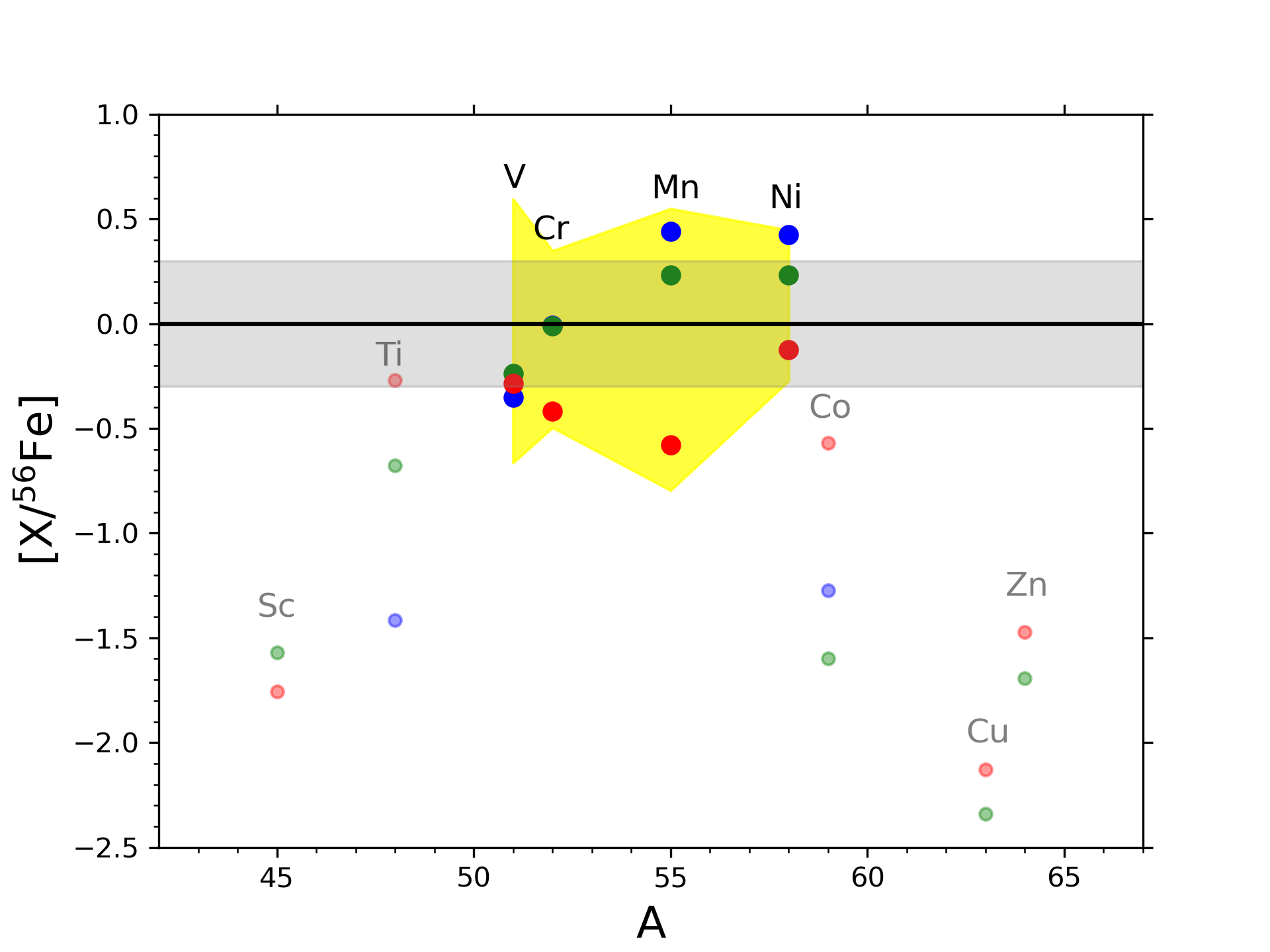}
    \caption{Solar metallicity [X/$^{56}$Fe] SN Ia yields of most abundant stable isotope for each Fe-peak element.  The yellow area cover the values spanned in the model tested for V, Cr, Mn and Ni. Blue, green and red points indicate the fractions for the PTD \citetalias{Leung18}, DDT \citetalias{Leung18} and DD \citetalias{Leung20} benchmark models, respectively. The grey area mark the region where the SN Ia yield is between two times and half of the solar value (solar normalisation is from \citealt{Asplund09}).}
    \label{f:yields}
\end{figure}

In this study we consider Fe-peak elements for which SN Ia production has a significant impact on the abundance patterns. These elements are vanadium, chromium, manganese and nickel.
For these four elements, in Figure \ref{f:yields} we show the range of values for [X/$^{56}$Fe] (where X is the most abundant isotope for an element) covered by the different SN Ia yields tested in this work at solar metallicity (yellow area). In this Figure, we see that not only the SN Ia pollution is important for these elements, but also that we have yields spanning even more than an order of magnitude.\\
We note also that other Fe-peak elements generally show lower [X/$^{56}$Fe] fractions.  For most of them it is claimed a predominant CC-SN production (e.g. Cu, Co; see \citealt{Ernandes20}). Moreover, for some elements severe problems are encountered in reproducing the observed trends (e.g. Ti, Sc; see \citealt{Romano10,Prantzos18}). For these reasons, we decide to not consider other Fe-peak elements in our analysis.

As done with [$\alpha$/Fe] ratios, in this Section we now concentrate on the differences between different yield sets. The comparison with several literature abundances, either with local thermodynamic equilibrium (LTE) assumption or non-LTE (NLTE) corrected, will be shown later.

\subsubsection*{Vanadium}

\begin{figure}
    \centering
    \includegraphics[width=1.\columnwidth]{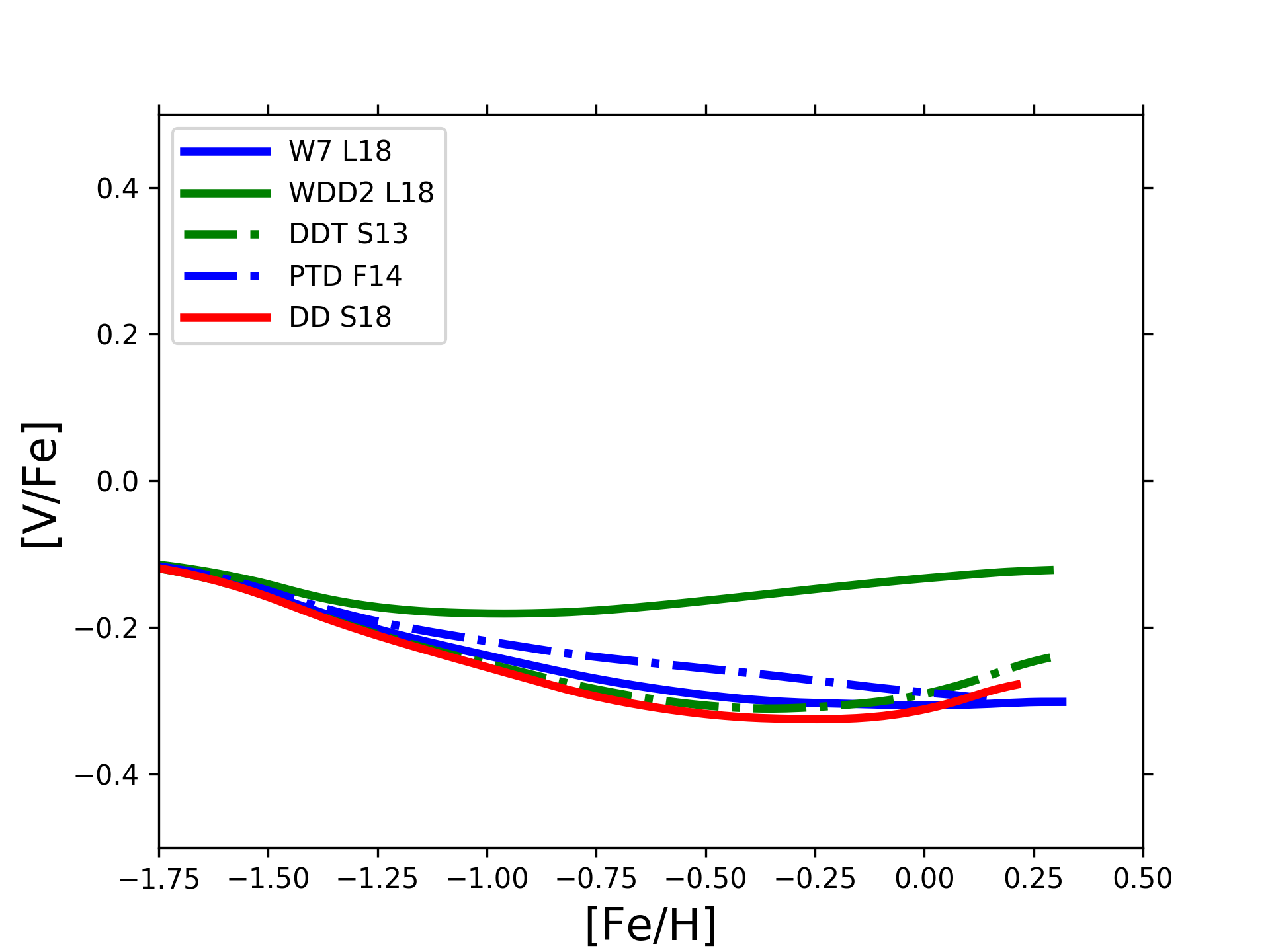}
    \caption{ Same of Figure \ref{f:MgFe}, but for [V/Fe]. Results are shown for W7 \citetalias{Leung18} (blue solid line), WDD2 \citetalias{Leung18} (green solid), DDT \citetalias{Seit13} (green dash-dotted), PTD \citetalias{Fink14} (blue dash-dotted) and DD \citetalias{Shen18} (red solid) yields.}
    \label{f:VFe}
\end{figure}

The contribution of different SN Ia yield sets to [V/Fe] vs. [Fe/H] is shown in Figure \ref{f:VFe}. All our benchmark models show a first decreasing and then an almost flat trend with metallicity, with only the  WDD2 model showing slightly larger [V/Fe] values. We see that the models shown in Figure \ref{f:VFe} exhibit subsolar [V/Fe] at solar metallicity: this can be also noted looking at the solar abundances for the benchmark yields shown in Table \ref{t:Fe_abundances} (in bold). However, part of this results may be due to the uncertain CC-SN contribution to V chemical enrichment  (we will come back later on this in Section \ref{ss:combo_yields}).

\begin{figure}
    \centering
    \includegraphics[width=1.\columnwidth]{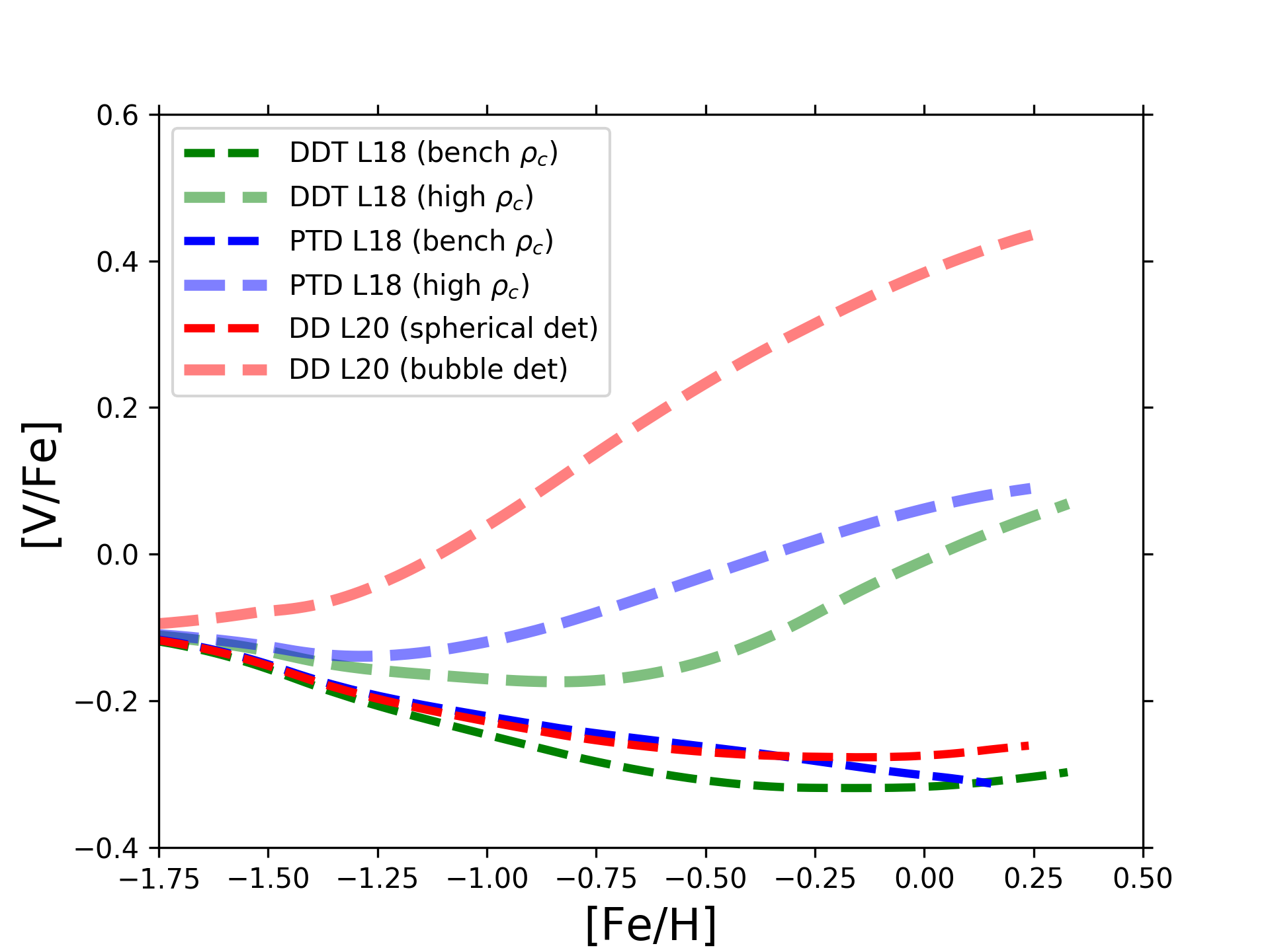}
    \caption{Variations in [V/Fe] vs. [Fe/H] diagram by adopting SN Ia yields with different initial conditions from the benchmark models. Results are shown for DDT \citetalias{Leung18} high density model (shaded green thick dashed line), PTD \citetalias{Leung18} high density model (shaded blue thick dashed) and DD \citetalias{Leung20} bubble detonation model (shaded red thick dashed). Results for the correspondent benchmark yields are shown with the usual color code.}
    \label{f:VFe_2}
\end{figure}

In Figure \ref{f:VFe_2} are plotted the DDT \citetalias{Leung18} model with larger WD central density ($5\times10^9$ gr cm$^{-3}$) than the benchmark model, the PTD \citetalias{Leung18} model also with larger central density and the sub-$M_{ch}$ \citetalias{Leung20} model with bubble He detonation trigger (i.e. with aspherical detonation pattern).\\
As we can see, not all the SN Ia yields tested in this work have subsolar [V/Fe]. Rather, highly supersolar [V/Fe] (by $\simeq0.4$ dex at solar metallicity, see Table \ref{t:Fe_abundances}) are obtained adopting sub-$M_{ch}$ yields with an aspherical He detonation. In fact, spherical symmetry in the He detonation pattern tends to produce less V. This is also a common feature with other light Fe-peak elements (e.g. Cr).
For what concerns the high density \citetalias{Leung18} DDT and PTD models plotted in Figure \ref{f:VFe_2} (and also for \citetalias{Seit13} and \citetalias{Fink14} ones, even if in much less extent) we see a larger V production than in the benchmark models, although not as much as for sub-Chandrasekar yields with bubble He detonation. The larger V production is caused by the much lager zone incinerated by deflagration in high density WD models.

\subsubsection*{Chromium}
\begin{figure}
    \centering
    \includegraphics[width=1.\columnwidth]{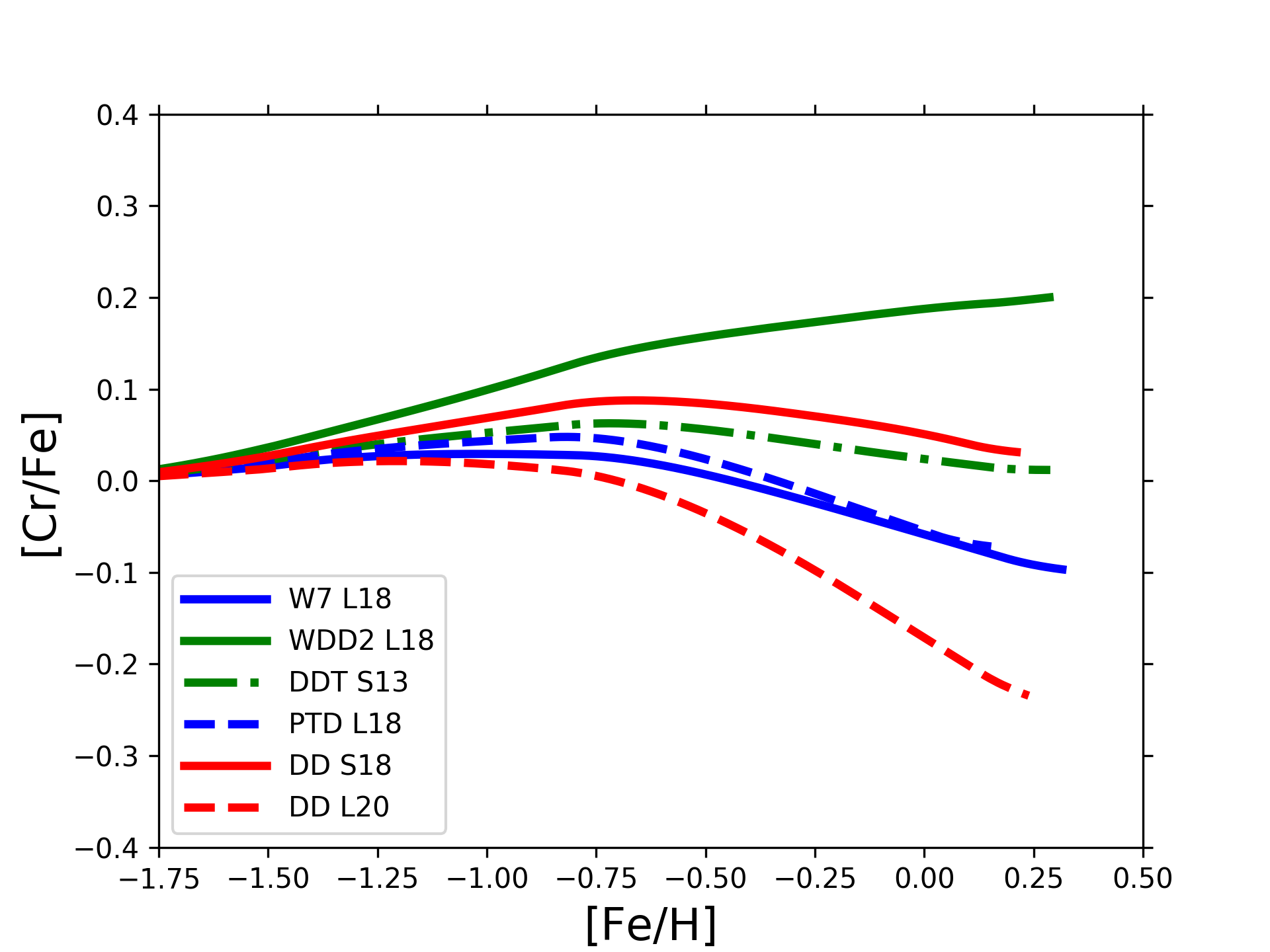}
    \caption{Same of Figure \ref{f:MgFe}, but for [Cr/Fe]. Results are shown for W7 \citetalias{Leung18} (blue solid line), WDD2 \citetalias{Leung18} (green solid), DDT \citetalias{Seit13} (green dash-dotted), PTD \citetalias{Leung18} (blue dashed), DD \citetalias{Shen18} (red solid) and DD \citetalias{Leung20} (red dashed) yields.}
    \label{f:CrFe}
\end{figure}

In Figure \ref{f:CrFe} we plot the model results for [Cr/Fe] vs. [Fe/H]. Our model adopting the benchmark DDT \citetalias{Seit13} yields shows a nearly flat trend up to supersolar metallicity. The same happens for the 1M$_\odot$, sub-$M_{ch}$ yields of \citet{Shen18}. Near-solar but slightly decreasing [Cr/Fe] values with metallicity are obtained by DDT \citetalias{Leung18} benchmark model, the PTD models by \citet{Fink14} and the PTD \citetalias{Leung18} benchmark model, as well as for the W7 model.
However, we remind that the result coming from W7 yields (also in the updated \citealt{Leung18} versions) is influenced by a severe overestimation of the $^{54}$Cr isotope. In fact, the models predict that almost 10$\%$ of the solar Cr is in the form of $^{54}$Cr, a factor 4 more than required to reproduce solar abundance ratio of Cr isotopes (\citealt{Bergemann10Cr}). A general underestimation of the $^{54}$Cr/$^{52}$Cr ratio is instead seen for the sub-$M_{ch}$ models tested.\\
For the other yield sets shown in Figure \ref{f:CrFe}, the [Cr/Fe] ratio tends to be overestimated (WDD2 \citetalias{Leung18}) or underestimated (sub-$M_{ch}$, DD \citetalias{Leung20} with spherical detonation) at solar metallicity. This can be also seen by looking at Table \ref{t:Fe_abundances}.

\begin{figure}
    \centering
    \includegraphics[width=1.\columnwidth]{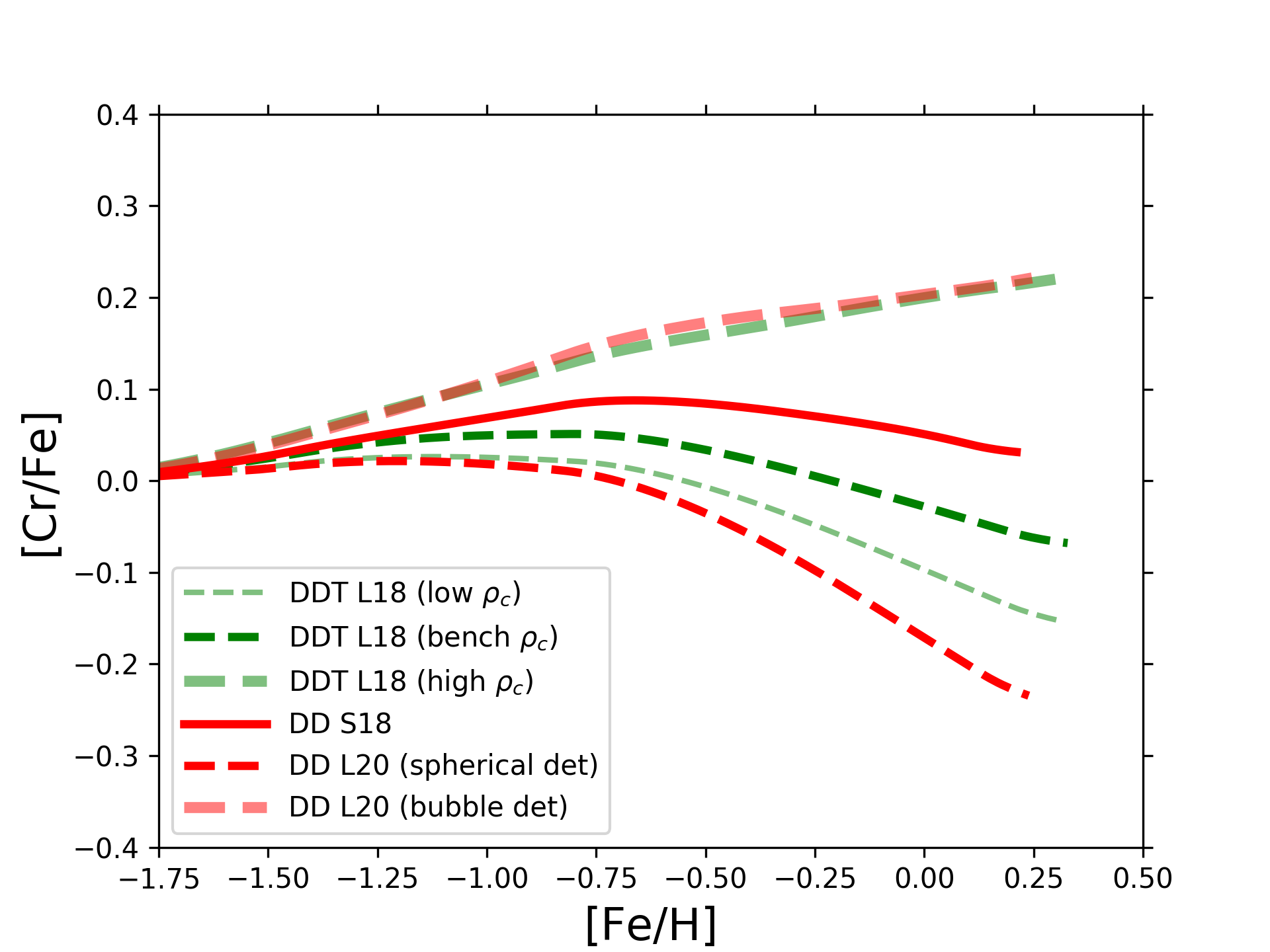}
    \caption{Variations in [Cr/Fe] vs. [Fe/H] diagram by adopting SN Ia yields with different initial conditions from the benchmark models. Results are shown for DDT \citetalias{Leung18} low density model (shaded green thin dashed line), DDT \citetalias{Leung18} high density model (shaded green thick dashed) and DD \citetalias{Leung20} bubble detonation model (shaded red thick dashed). Results for the correspondent benchmark yields are shown with the usual color code. DD \citetalias{Shen18} model (red solid line) is also plotted for comparison with DD \citetalias{Leung20} benchmark model.}
    \label{f:CrFe_2}
\end{figure}

However, as for V, the WD initial conditions for the simulated SN Ia have an important impact on the nucleosynthesis. In Figure \ref{f:CrFe_2} we show the model results for DDT \citetalias{Leung18} models with different WD central density, and a sub-$M_{ch}$, DD \citetalias{Leung20} model with different He detonation pattern (bubble configuration).\\
As it happens for V, we see an increment in [Cr/Fe] ratio due to the overproduction of Cr at higher densities (by $\sim$0.25 dex, see also Table \ref{t:Fe_abundances}). Conversely, a lower Cr production is obtained at lower densities.  We highlight that the same behaviour with WD central density is obtained with pure deflagration models. 
Concerning the sub-$M_{ch}$ DD models, we see that the Cr production is favoured for an aspherical detonation pattern (as for the others light Fe-peak elements).\\
However, we note that the behaviour of [Cr/Fe] with the WD central density or initial detonation pattern are not the same for different studies.  Models with \citet{Shen18} and \citet{Leung20} (benchmark) sub-$M_{ch}$ yields show a variation of $\gtrsim 0.2$dex in their solar Cr abundances despite of the same symmetric detonation structure (see also Figure 39 of \citealt{Leung20}). At the same time, WD central density variations in \citet{Seit13} (and partly in \citealt{Fink14}) do not produce a significant [Cr/Fe] enhancement/decrease as in \citet{Leung18}. For these latter cases, the slightly different initial conditions 
can explain the different behaviour of the yields.

\subsubsection*{Manganese}
Manganese has been extensively studied in astronomical literature. However, the contribution to its production from different types of SNe is still uncertain (e.g. \citealt{Seit13b,Eitner20}).

\begin{figure}
    \centering
    \includegraphics[width=1.\columnwidth]{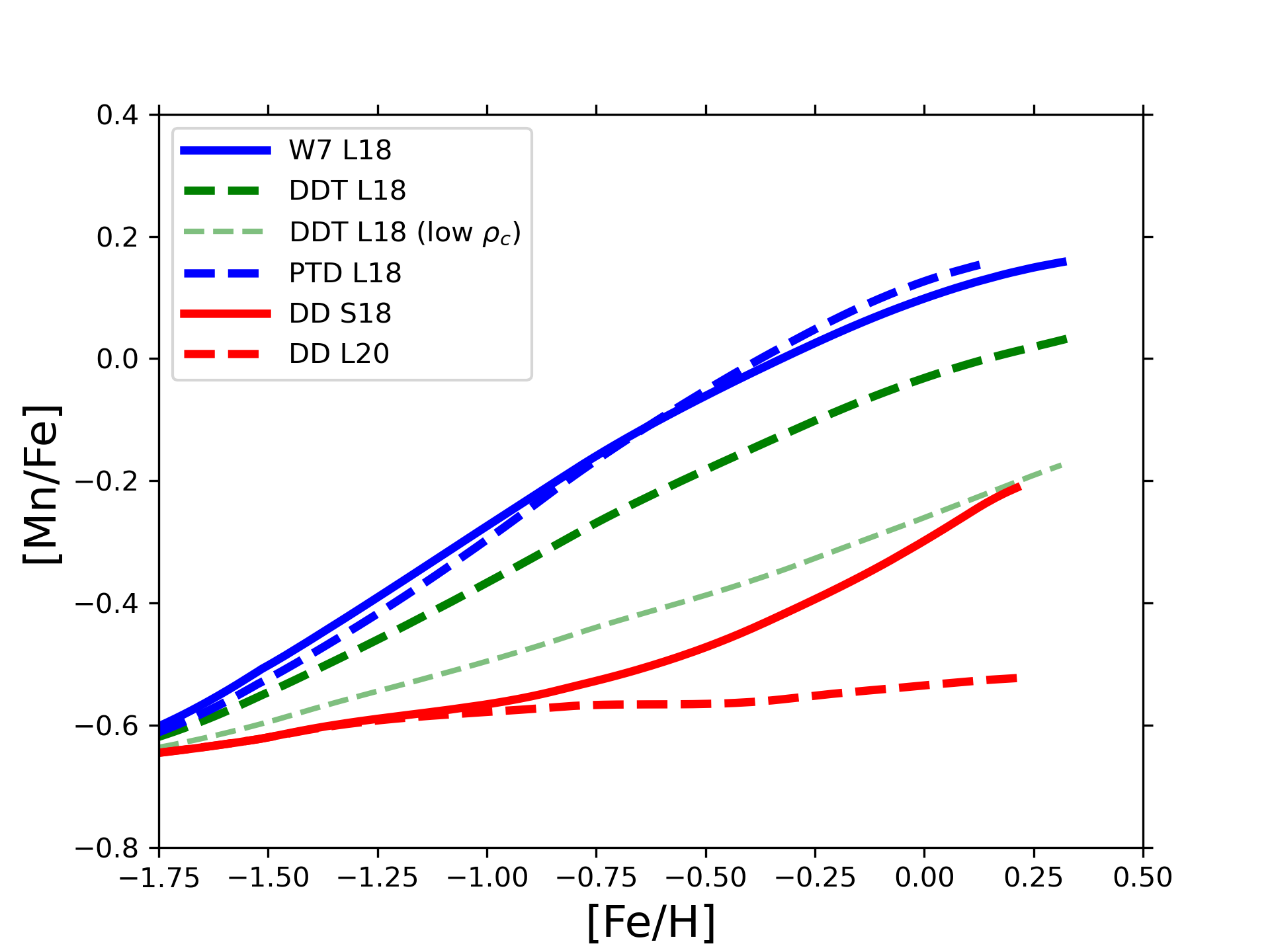}
    \caption{Same of Figure \ref{f:MgFe}, but for [Mn/Fe]. Results are shown for W7 \citetalias{Leung18} (blue solid line), DDT \citetalias{Leung18} (green dashed), DDT \citetalias{Leung18} low density model (shaded green thin dashed),  PTD \citetalias{Leung18} (blue dashed), DD \citetalias{Shen18} (red solid) and DD \citetalias{Leung20} (red dashed) yields.}
    \label{f:MnFe}
\end{figure}

As we can see from Figure \ref{f:MnFe}, multi-D benchmark yields adopting DDT and PTD explosion mechanisms show a qualitatively similar behaviour, with [Mn/Fe] jumping from highly subsolar values to solar ones. This is caused by the fact that Mn is mainly synthesised during the deflagration phase (see Figure 7 of \citealt{Leung18}).\\
The same of multi-D models happens for W7 and WDD2 yields, where a similar jump to solar values is seen.  
For what concerns sub-$M_{ch}$ models, [Mn/Fe] at solar metallicities is lowered by at least $\sim0.25$ dex (see also Table \ref{t:Fe_abundances}). A similar offset is expected, since the limited electron capture in sub-Chandrasekar WD explosions (\citealt{Leung20}). An offset is also present between different sub-$M_{ch}$ yields, with \citet{Shen18} showing larger [Mn/Fe] values relative to \citet{Leung20} (especially when adopting the benchmark/spherical He detonation configuration).

For what concerns DDT \citetalias{Leung18} models, we see in Figure \ref{f:MnFe} that a lower WD central density (i.e.  a WD mass of 1.33M$_\odot$) leads to subsolar [Mn/Fe] at solar metallicity (see also Table \ref{t:Fe_abundances}). A similar decrease of [Mn/Fe] with central density is found for PTD \citetalias{Leung18} models, which however still show larger abundance ratios due to the lack in Fe production. This "deficit" in Mn pollution (roughly 0.2 dex) relative to the benchmark models is due to the more massive zones incinerated by detonation instead of deflagration. As for Cr, this behaviour with WD central density does not hold if we consider \citet{Seit13} yields due to the different conditions in the simulated SNe.

\subsubsection*{Nickel}
Nickel is another key element to understand the contribution to the chemical enrichment of the different SN Ia channels (e.g. \citealt{Kirby19}). Ni was also one of the main concerns for the W7 model by \citet{Iwa99} in chemical evolution, predicting Ni/Fe ratios much higher than what observed at [Fe/H]$\gtrsim$-1 (\citealt{Koba06,Romano10,Prantzos18}). 

\begin{figure}
    \centering
    \includegraphics[width=1.\columnwidth]{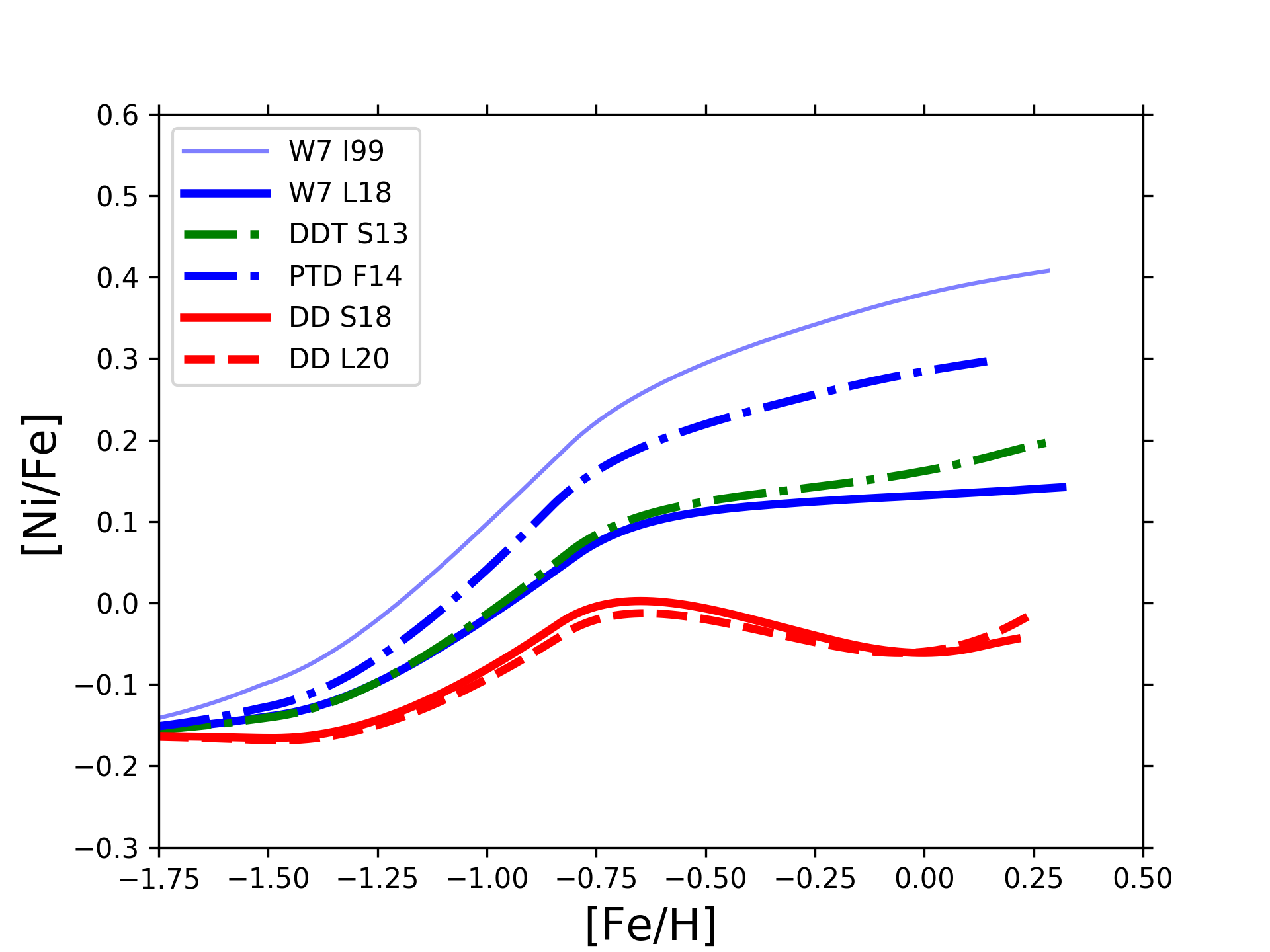}
    \caption{Same of Figure \ref{f:MgFe}, but for [Ni/Fe]. Results are shown for W7 \citetalias{Leung18} (blue solid line), WDD2 \citetalias{Leung18} (green solid), DDT \citetalias{Seit13} (green dash-dotted), PTD \citetalias{Fink14} (blue dash-dotted) and DD \citetalias{Shen18} (red solid) yields.  We also show the result for W7 \citet{Iwa99} (\citetalias{Iwa99}) yields (shaded blue thin solid line) for direct comparison with the updated \citet{Leung18} version.}
    \label{f:NiFe}
\end{figure}

\begin{table*}
    \centering
    \caption{Combination of near-$M_{ch}$ and sub-$M_{ch}$ SN Ia progenitors tested in this paper. The rightmost column shows the suffix placed to indicate variations from near-$M_{ch}$ DDT \citetalias{Leung18} and sub-$M_{ch}$ DD \citetalias{Leung20} benchmark models.}
    \begin{tabular}{c | c  c}
        \hline
        \hline
        Model  &  Combination  &  Variation suffix \\
        \hline
        100n & 100\% near-$M_{ch}$ DDT & \\[0.05cm]
        75n-25s & 75\% near-$M_{ch}$ DDT - 25\% sub-$M_{ch}$ DD & l=low $\rho_c$ DDT \citetalias{Leung18} model\\[0.05cm]
        50n-50s & 50\% near-$M_{ch}$ DDT - 50\% sub-$M_{ch}$ DD & h=high $\rho_c$ DDT \citetalias{Leung18} model\\[0.05cm]
        25n-75s & 25\% near-$M_{ch}$ DDT - 75\% sub-$M_{ch}$ DD & b=bubble detonation DD \citetalias{Leung20} model\\[0.05cm]
		100s & 100\% sub-$M_{ch}$ DD & \\
        \hline
        \hline
    \end{tabular}
    \label{t:combo_models}
\end{table*}

Looking at Figure \ref{f:NiFe}, we note that pure deflagration yields (\citealt{Fink14,Leung18}) lead to a rapid increase in [Ni/Fe] between [Fe/H]$\sim$-1.25 and [Fe/H]$\sim$-0.75. The enhancement in Ni is less prominent for the benchmark DDT models. We also note that the model with the W7 \citetalias{Leung18} yields (with updated nuclear reaction network) shows values similar to the models adopting DDT yields, with an offset of $\sim0.2$ dex at solar metallicity relative to the chemical evolution model adopting the old W7 \citet{Iwa99} version. The difference can be explained by the lower electron capture rates in the updated models (\citealt{Koba20}).\\
On the contrary, sub-$M_{ch}$ yields (\citealt{Shen18,Leung20}) are not able to produce supersolar [Ni/Fe] in the chemical evolution pattern, placing the [Ni/Fe] at subsolar values at [Fe/H]=0. \\
Similarly to Mn, we highlight that DDT and PTD \citetalias{Leung18} low density models show lower Ni pollution. In this case, however, this fact holds also for the models by \citet{Seit13} and \citet{Fink14}, as can be seen in Table \ref{t:Fe_abundances} for the predicted solar abundances.

The different behaviour seen in the models is due to fact that the Ni yield is strongly dependent on the electron fraction $Y_e$ (and hence on the efficiency of electron capture), as it happens for Mn. As said above for the W7 model, Ni production is disfavoured for higher $Y_e$ (approaching 0.5), a condition that is found in sub-$M_{ch}$ models. This dependence on $Y_e$ also allows the DDT \citetalias{Leung18} low density model to have a lower [Ni/Fe] yield.
In the case of multi-D PTD models instead, the [Ni/Fe] yield is enhanced relative to DDT models, since $^{58}$Ni tend to be overproduced because of the lower $^{56}$Fe mass.

\subsection{Combination of different sets}
\label{ss:combo_yields}

Since we have tested the nucleosynthesis of different SN Ia explosion mechanisms, it is of great interest to study if a combination of them can improve the agreement with the observed abundances.  Lightcurve and abundance observations of individual SNe Ia show characteristics typical of different subclasses (e.g. \citealt{Kirby19,DeLosReyes20}), suggesting that SNe Ia explode through different mechanisms.  Moreover,  studies on SN Ia rate (e.g. \citealt{Maoz14}) and on WDs population in the solar vicinity (Gentile Fusillo,  private communication) advocate similar conclusions, with some fraction of sub-$M_{ch}$ progenitors that seems necessary to explain the  inferred features. \\
Some recent works aimed at assessing the role of different SN Ia progenitors in the framework of chemical evolution but most of them focused only at single elements, such as manganese (e.g. \citealt{Seit13b,Eitner20}), or looked at other systems, mostly dwarf MW satellites (e.g. \citealt{Koba15,Cescutti17}).\\

Here we concentrate our analysis on Fe-peak elements, and in particular to V, Mn, Cr and Ni, which show the most significant variations between different SN Ia models (remember Figure \ref{f:yields}).\\
The different progenitor combinations tested are shown in Table \ref{t:combo_models}, where in the first column we list the model names for the specific progenitor combinations, indicated in the second column. 
As we can see in Table \ref{t:combo_models},  we look at the combinations of near-$M_{ch}$, DDT yields with sub-$M_{ch}$, DD yields.  Multi-D near-$M_{ch}$ PTD models are likely to be representative of SNe Iax,  which very likely leave a remnant, have weak explosions and which low ejecta mass may not be important for chemical enrichment compared to other explosion models (e.g. \citealt{Leung18,Koba20b}).  \\
For this analysis, we adopt the yields by \citet{Leung18} (for DDT models) and \citet{Leung20} (DD models), due to their deep exploration of the parameter space. 
In fact, in \citet{Leung18} the WD central density/mass was varied in different DDT models. Such variations are consistent, for example, with those allowed for the SD scenario, where the WD central density/mass are strictly connected to the accretion rate from the non-degenerate companion star (e.g. \citealt{Nomoto94,Leung18}). At the same time, in \citet{Leung20} DD models multiple types of detonation-triggers were investigated. These models leave abundance patterns of most characteristic iron-peak elements for sub-$M_{ch}$ SNe Ia (Mn, Ni) which are compatible with the sub-$M_{ch}$ scenario, but also show very different patterns for other Fe-peak elements.
The adoption of these different yield sets is indicated with a suffix in the model name, for which a list in the third column of Table \ref{t:combo_models} is shown.

At variance with Section \ref{ss:single_yields}, here we apply the yields to the revised two-infall model of \citet{Palla20b}, which best describes the abundances, as well as stellar ages, in the solar neighbourhood. The model is able to explain recent APOGEE (\citealt{Hayden15}) and APOKASC (\citealt{Silva18}) data (see \citealt{Palla20b}).  The parameters of the models also allow to reproduce a present-day star formation rate (SFR) density in line with that observed in the solar vicinity. In fact, the predicted value for SFR is:
\begin{equation*}
\Sigma_{SFR,predicted}=4.34 $ M$_\odot $ pc$^{-2} $ yr$^{-1},
\end{equation*}
which is in good agreement with the range of values given by \citet{Prantzos18}:
\begin{equation*}
\Sigma_{SFR,observed}=2-5$ M$_\odot $ pc$^{-2} $ yr$^{-1}.
\end{equation*}


Our model results are compared with several literature abundance ratios. We adopt as for $\alpha$-elements (Figure \ref{f:alphaFe_calibr}) the abundances of \citet{Chen00} (for Mn we adopt a subsample of \citealt{Nissen00} which correspond to \citealt{Chen00} stars); \citet{Adibekyan12} and \citet{Bensby14} for moderate to high [Fe/H] stars. For metal poor halo-thick disc stars we use instead the measurements from \citet{Lai08} and \citet{Ou20}.
In addition, we consider non-LTE (NLTE) corrected measurements from \citet{Bergemann10Cr} and \citet{Eitner20}. In fact, for several Fe-peak elements, NLTE effects in neutral lines were found to be large, especially in the metal-poor regime. This may cause a substantial change in the chemical evolution picture for the element.\\

In order to asses also quantitatively the best models in reproducing the abundance data trend, we perform a statistical test.  In this way, we determine the best yield combinations to reproduce each abundance diagram (V, Cr, Mn, Ni) as well as the overall best model for Fe-peak elements tested here.\\
In particular, we run a quasi-$\chi^2$ diagnostic, defined in this way:
\begin{equation}
    \chi^2=\frac{1}{N}\sum_{n=1}^N \, \sum_{i=1}^I \, \bigg(\frac{X_{dat;n,i}-X_{mod;n,i}}{\sigma_{dat;n,i}}\bigg)^2,
    \label{e:chi_quadro}
\end{equation}
where the sum is calculated over the $N$ data points in the abundance diagram (index $n$) and the observables [X/Fe] and [Fe/H] (index $i$). Since there is more than one [X/Fe] values for a certain [Fe/H] (see \citealt{Palla20b} and subsequent Figures), it becomes ambiguous to associate an observed data point in the abundance space to a point on the model track. As in \citet{Spitoni20}, we associate a data point to the closest point in the curve. Given a data point $X_{dat;n,i}$, this is done by defining this function:
\begin{multline}
    S_n= \min_j \Bigg\{ \sqrt{\sum_{i=1}^I \, \bigg( \frac{X_{dat;n,i}-X_{mod;n,i,j}}{\sigma_{dat;n,i}} \bigg)^2 } \Bigg\} = \\ = \sqrt{\sum_{i=1}^I \, \bigg( \frac{X_{dat;n,i}-X_{mod;n,i,j'}}{\sigma_{dat;n,i}} \bigg)^2 },
    \label{e:min_distance}
\end{multline}
where $j$ are the runs over the different points on the curve. Hence, the closest point on the curve is $X_{mod;n,i} = X_{mod;n,i,j'}$.\\
Having computed the quasi-$\chi^2$ for each abundance diagram, the results are then summed to obtain the overall data-model agreement.

Since the uncertainties at low metallicity for V and Mn (see later), we repeat the process for two times, one time adopting standard CC-SN yields (from \citealt{Koba06}) and another time using modified CC-SN yield values that account for the observed behaviour at low metallicity. For the same reason, we decide to consider in the calculation only stars from \citet{Chen00} (\citealt{Nissen00}); \citet{Adibekyan12} and \citet{Bensby14} samples.

\subsubsection*{Vanadium}
As we have seen in Figures \ref{f:VFe} and \ref{f:VFe_2}, the largest variations in the chemical evolution of vanadium are not driven by the different type of explosion mechanisms (i.e. PTD, DDT, DD). The main reason of the spread of the models in [V/Fe] is due to the initial condition of the exploding WD. This is particularly evident for sub-$M_{ch}$ models, where different He detonation patterns can produce variations up to $\sim0.6$ dex. In this way, we can use [V/Fe] as an indicator of the asphericity in He detonation in sub-$M_{ch}$ progenitors (\citealt{Leung20}).

\begin{figure}
    \centering
    \includegraphics[width=1.\columnwidth]{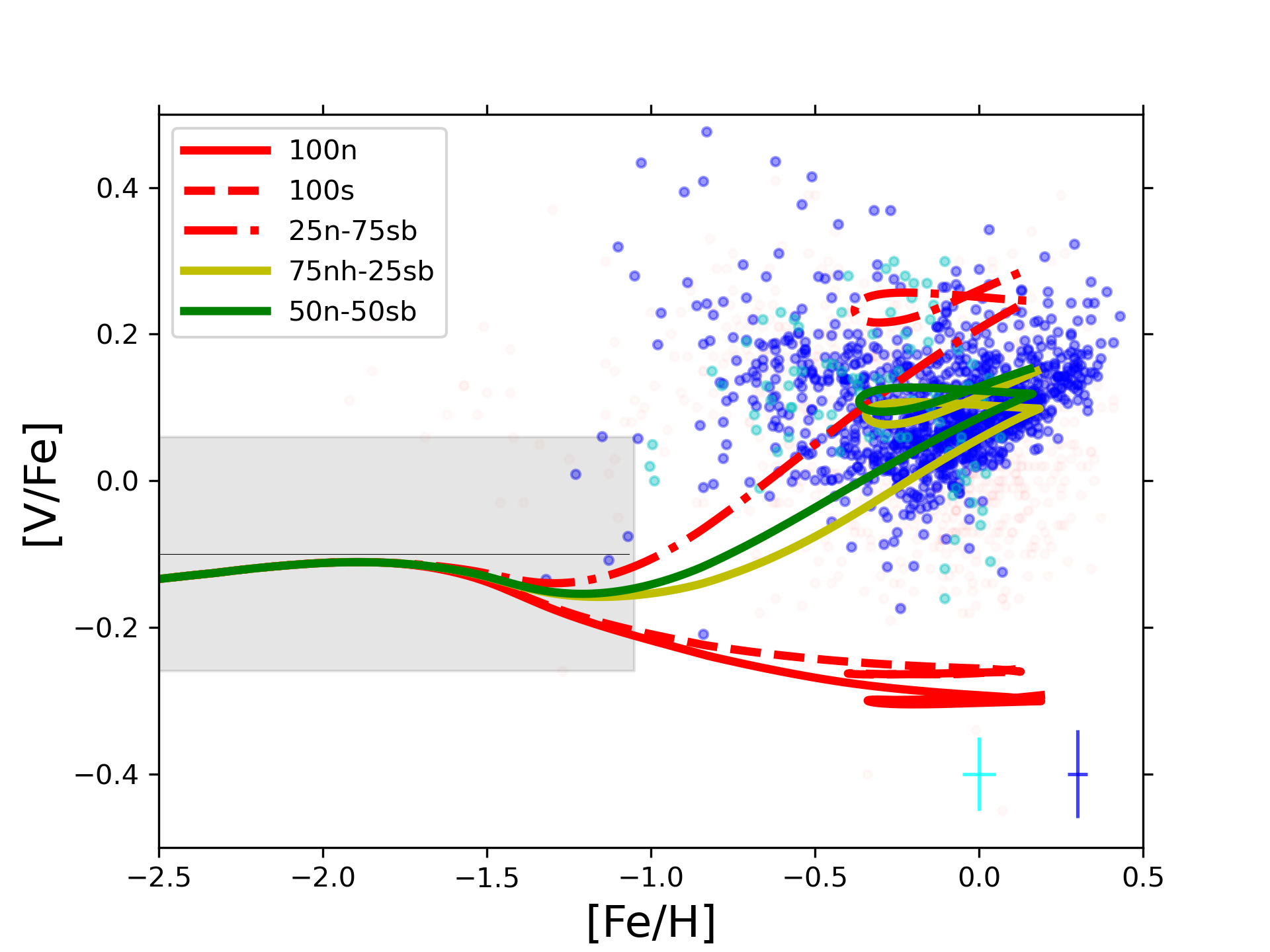}
    \includegraphics[width=1.\columnwidth]{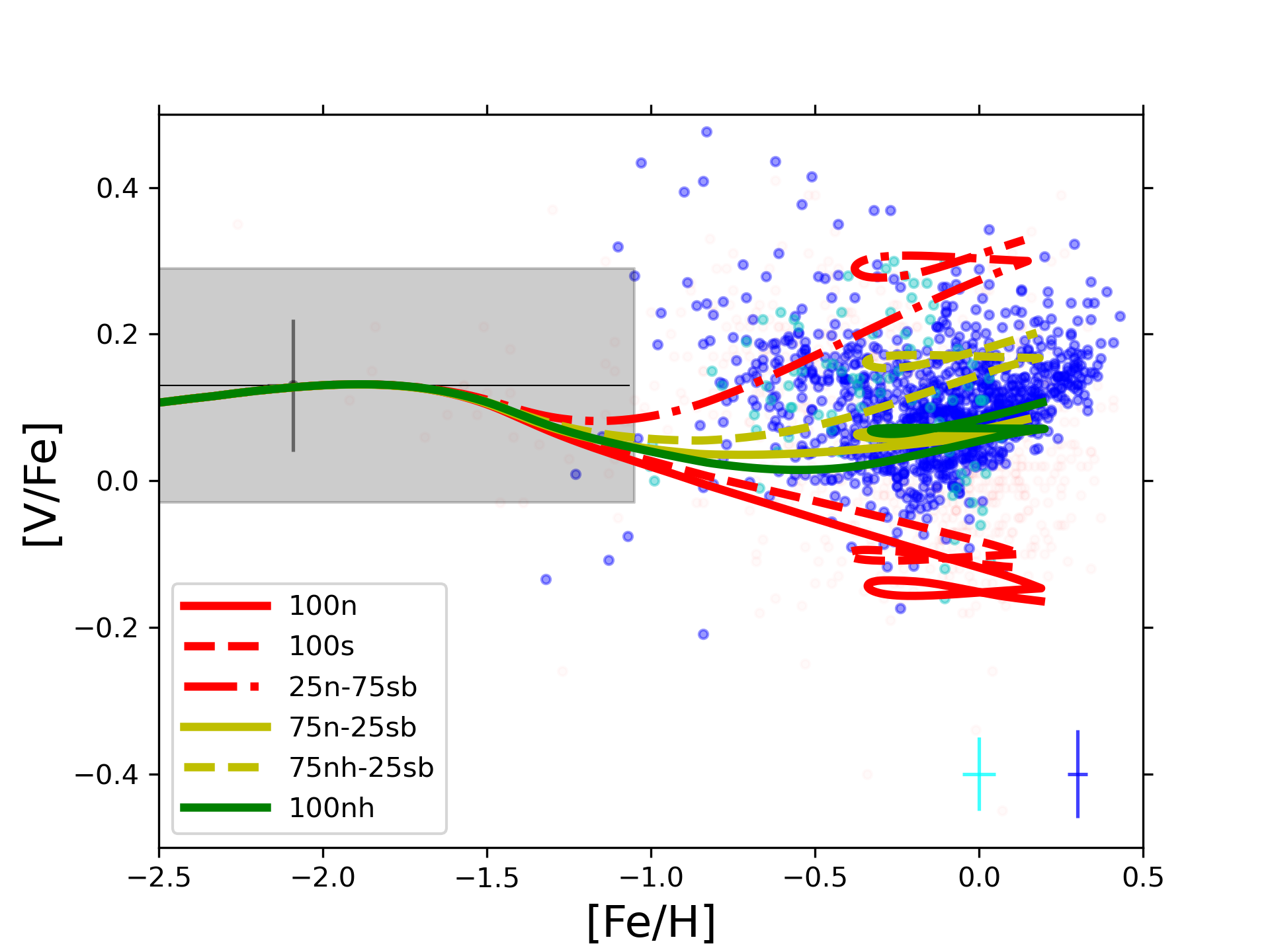}
    \caption{[V/Fe] vs. [Fe/H] ratios predicted by our two-infall chemical evolution model adopting different combinations of different SN Ia progenitor yield sets (see Table \ref{t:combo_models}). Green lines stand for the best models obtained by the statistical test; yellow lines indicate other models able to explain the observed data trend; red lines represent models with bad agreement with the observations. Upper panel: results for model with standard CC-SN yields from \citet{Koba06}. Lower panel: results for model with vanadium CC-SN yields multiplied by a 1.75 factor. 
    Both panels show VI LTE data are \citet{Chen00} (cyan points) and \citet{Adibekyan12} (blue points).  Cyan and blue errorbars indicate the typical uncertainties in \citet{Chen00} and \citet{Adibekyan12} samples. Upper panel shows also VI LTE \citet{Ou20} sample average value with correspondent rms (light grey line with shaded area). VII LTE data from \citet{Lai08} (grey points) and \citet{Ou20} sample average value with correspondent rms (grey line with shaded area) are shown in the lower panel.}
    \label{f:VFe_combo}
\end{figure}

In Figure \ref{f:VFe_combo} we show the results for different combinations of SN Ia progenitors. In particular, in the upper panel we look at the V evolution adopting standard CC-SN yield (\citealt{Koba06}), which agree well with [VI/Fe] measurements at low-metallicity (\citealt{Ou20}, light grey area). Since vanadium abundance from VI lines is probably underestimated by at least $\sim 0.1$ dex (\citealt{Scott15}), in Figure \ref{f:VFe_combo} lower panel we provide also the results for models with vanadium CC-SNe yields multiplied by a 1.75 factor. This change provide better agreement with low metallicity [V/Fe] inferred from VII lines (\citealt{Ou20}, grey area), that should be less affected by NLTE effects (V should be predominantly singly-ionised in stellar atmosphere due to its low ionisation energy, \citealt{Scott15}). Unfortunately, however, no NLTE measurements are at disposal for V to confirm what said above.\\
In both panels, we note the characteristic loop feature of the models of \citet{Palla20b}. This behaviour is the consequence of a delayed second infall, which dilutes the local ISM of primordial gas lowering the [Fe/H] ratio and leaving the [X/Fe] unchanged. The metal abundance is then restored thanks to the star formation (see also \citealt{Spitoni19,Palla20b}).\\
Looking at both upper and lower panels of Figure \ref{f:VFe_combo}, we see that both near-$M_{ch}$ (100n) and sub-$M_{ch}$ (100s) benchmark yields are not able to reproduce the bulk of the data. At the same time, a large fraction of sub-$M_{ch}$ progenitors with aspherical detonation trigger (i.e. bubble detonation) is unlikely, since it raises too much the V abundance. Therefore, models adopting moderate fraction (up to 0.5) of this latter progenitor subclass assure a good agreement with data. In particular, this can be seen for the upper panel, where the best model (green line) exhibit a sub-$M_{ch}$, bubble detonating SN Ia fraction of 0.5. We also note that a significant contribution from near-$M_{ch}$ SN Ia progenitors with high WD central density (in particular when we adopt \citetalias{Leung18} yields) increases the [V/Fe] ratio at high metallicity.  In fact, the near-$M_{ch}$, high central density set (100nh) results the best to explain the [V/Fe] behaviour when applied to a model with high V production by CC-SN (lower panel of Figure \ref{f:VFe_combo}).  However, a contribution from sub-$M_{ch}$ progenitors with aspherical detonation cannot be left out also in this case. 

Nonetheless, we have to take these results with caution.  The uncertainties on V abundances at low metallicity do not well constrain the V enrichment from CC-SNe, providing limits on the analysis of SN Ia contribution.\\

\subsubsection*{Chromium}

In Section \ref{ss:single_yields} we saw that the initial conditions for the simulated SN Ia have an important impact on the nucleosynthesis of chromium (Figure \ref{f:CrFe_2}). For this reason, we can use Cr as an indicator of the detonation pattern for sub-$M_{ch}$ progenitors (\citealt{Leung20}) and of the WD central density (and hence WD mass) for near-$M_{ch}$ progenitors (\citealt{Leung18}).

Concerning [Cr/Fe] observations, we adopt abundances inferred from LTE CrI lines only for moderate to high metallicities, where NLTE effects are found to be less prominent ($\sim0.1$ dex, \citealt{Prochaska00}). In fact, LTE calculations have revealed severe problems in determining the Cr abundance in metal poor stars, due to the very large offset between values inferred from CrI and CrII (up to 0.5 dex, e.g. \citealt{Gratton03,Lai08}). For this reason, for [Fe/H]$<-1.5$ dex, we consider only CrI NLTE abundances or CrII LTE abundances (whose LTE-NLTE offset is much lower than for CrI, \citealt{Bergemann10Cr}).\\
Such data selection provides a good agreement with our chemical evolution model at low metallicity, indicating the reliability of the adopted CC-SN yields for Cr. 

\begin{figure}
    \centering
    \includegraphics[width=1.\columnwidth]{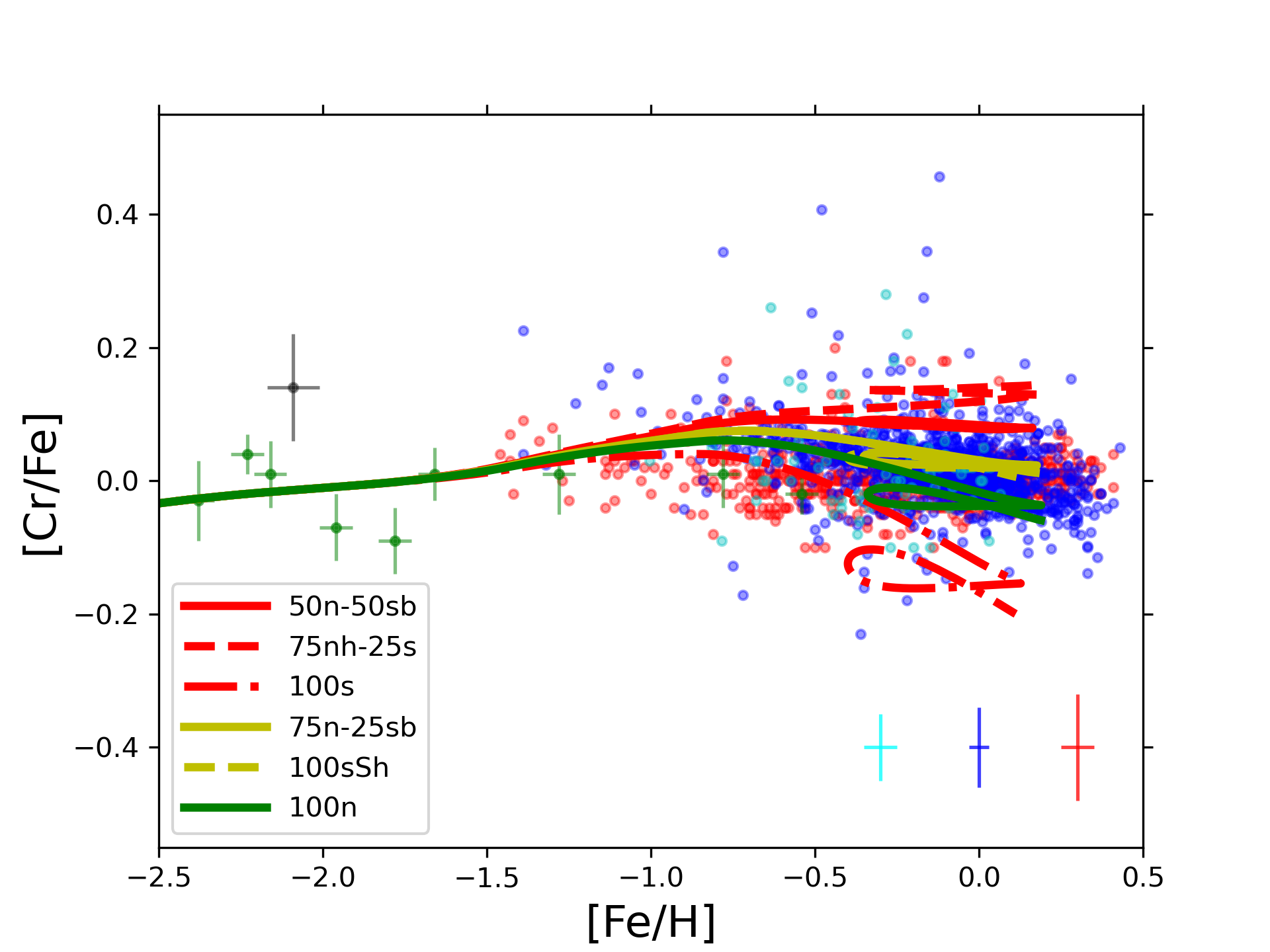}
    \caption{Same of Figure \ref{f:VFe_combo} but for [Cr/Fe]. CrI LTE data are from \citet{Chen00} (cyan points). Mediated CrI+CrII LTE data are from \citet{Bensby14} (red points). CrII LTE data are from \citet{Lai08} (grey points with errorbars) \citet{Adibekyan12} (blue points). Cr NLTE data are from \citet{Bergemann10Cr} (green points with errorbars).  Blue and red errorbars indicate the typical uncertainties in \citet{Adibekyan12} and \citet{Bensby14} samples.  Dashed yellow line (100sSh) adopt DD \citetalias{Shen18} yields for sub-$M_{ch}$ progenitors.}
    \label{f:CrFe_combo}
\end{figure}

Looking at Figure \ref{f:CrFe_combo}, we note that an important fraction ($> 0.25$) of sub-$M_{ch}$ progenitors with aspherical detonation is not favoured by the observational constraints (see red solid line). At the same time, a little but not negligible fraction of this progenitor class is likely contributing to the enrichment, since the low [Cr/Fe] yield for sub-$M_{ch}$, spherical detonation \citetalias{Leung20} models (see red dash-dotted line).  However, we must say that \citetalias{Shen18} sub-$M_{ch}$, spherical detonation model (indicated as $\lq$sSh' in this Section) produces a good agreement with the observations (remember Section \ref{ss:single_yields}).\\
Concerning near-$M_{ch}$ progenitors, the data do not favour the scenario in which a considerable fraction of them are high central density/mass WD.  Their presence does not allow to reproduce the observational trend, as we can note from Figure \ref{f:CrFe_combo} (red dashed line). On the contrary, the near-$M_{ch}$ benchmark model is in very good agreement with the data. In fact, this yield set results the best from the statistical test for [Cr/Fe] vs. [Fe/H]. \\
We have to say that these considerations about initial conditions for near-$M_{ch}$ WDs are valid in the case we are adopting \citet{Leung18} yields. In fact, if we use the yields of \citet{Seit13}, we do not see any particular variation in changing the WD density (remember Section \ref{ss:single_yields}).\\

\subsubsection*{Manganese}
We already saw in Figure \ref{f:MnFe} that we have substantially different [Mn/Fe] yields for different SN Ia explosion mechanisms. Higher [Mn/Fe] values are obtained using PTD yields relative to DDT ones, which in turn give much larger Mn production relative to sub-$M_{ch}$ DD explosions.

For these differences between different classes of models, Mn is considered the most promising element to identify the contributions to chemical enrichment of different SN Ia progenitor classes (e.g. \citealt{Cescutti17,Eitner20,DeLosReyes20}). However, the very different [Mn/Fe] values inferred over the whole metallicity range limit the conclusions of many of the previous works (e.g. \citealt{Seit13b}).  Up to recent years LTE abundances suggested that the [Mn/Fe] is highly sub-solar in low metallicity stars (e.g. \citealt{Bonifacio09}). However, in the last years NLTE studies show that Mn lines are severely affected by NLTE effects, suggesting a [Mn/Fe] closer to solar at low metallicity (\citealt{Bergemann08,Bergemann19,Eitner20}).

\begin{figure}
    \centering
    \includegraphics[width=1.\columnwidth]{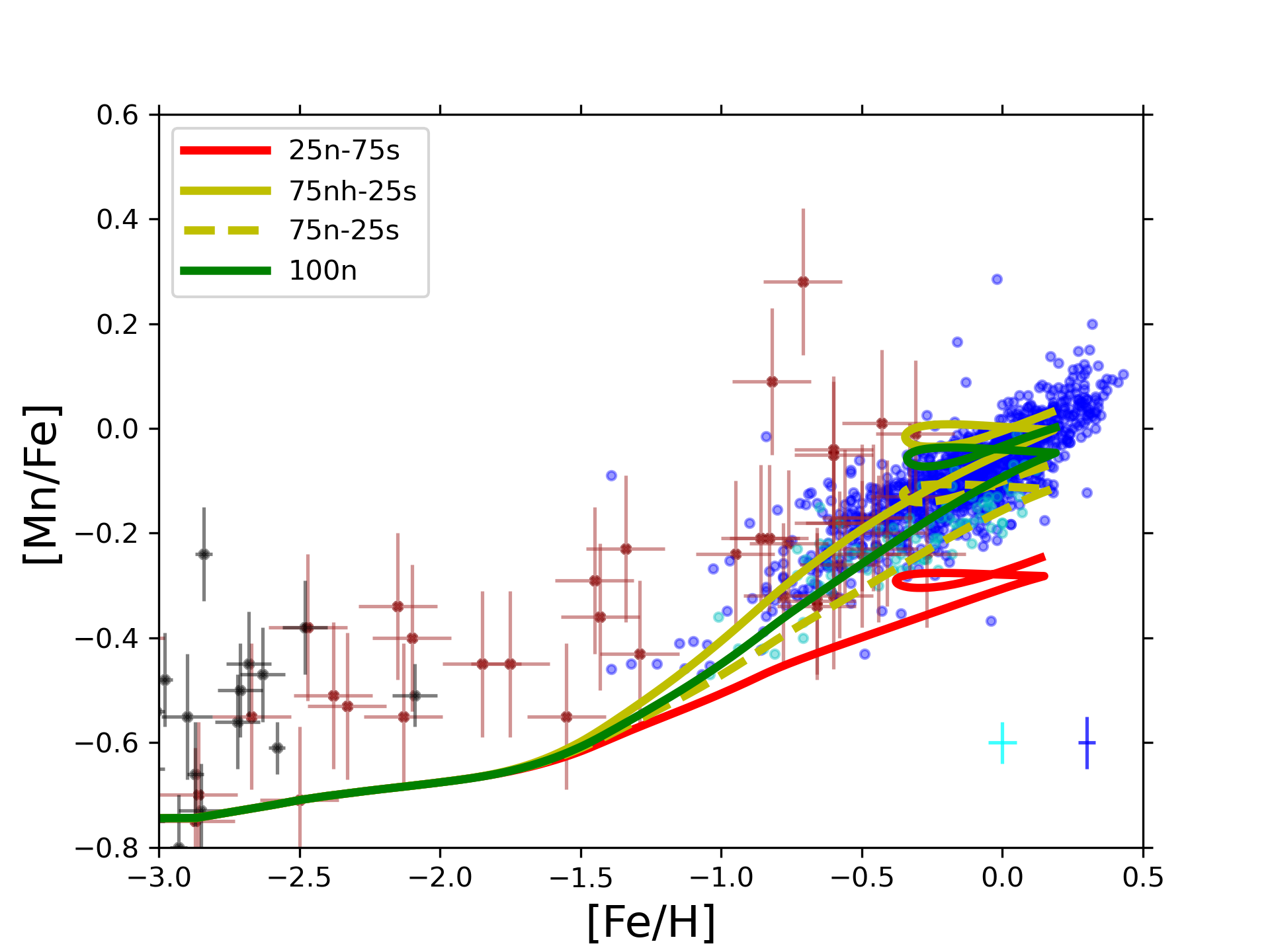}
    \includegraphics[width=1.\columnwidth]{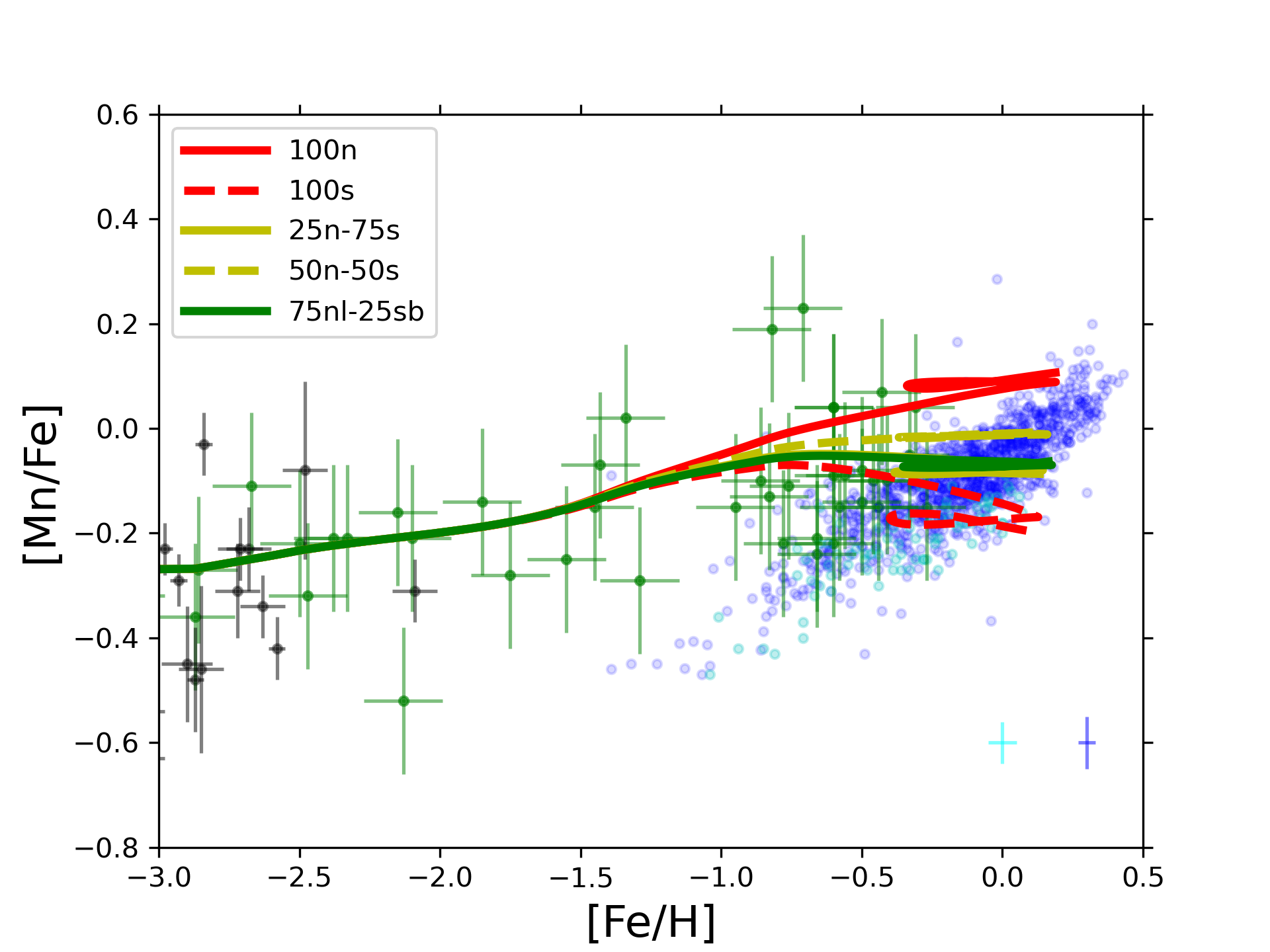}
    \caption{Same of Figure \ref{f:VFe_combo} but for [Mn/Fe]. Upper panel: results for model with standard CC-SN yields from \citet{Koba06}. Lower panel: results for model with manganese CC-SN yields multiplied by a 3 factor. 
    Both panels show Mn LTE data from \citet{Nissen00} (cyan points) and \citet{Adibekyan12} (blue points).  Cyan and blue errorbars indicate the typical uncertainties in \citet{Nissen00} and \citet{Adibekyan12} samples. Upper panel shows also Mn LTE data from \citet{Lai08} (grey crosses with errorbars) and \citet{Eitner20} (brown dots with errorbars). Lower panel shows instead MnII LTE data from \citet{Lai08} (grey points with errorbars) and Mn NLTE data from \citet{Eitner20} (green points with errorbars).  In this panel, a shaded effect is given to \citet{Nissen00} and \citet{Adibekyan12} data. }
    \label{f:MnFe_combo}
\end{figure}

In Figure \ref{f:MnFe_combo}, we consider the effect of combining near-$M_{ch}$ DDT and sub-$M_{ch}$ DD yields.  In the upper panel, we show the results of the two-infall model adopting standard CC-SN yields from the literature (\citealt{Koba06}). In this panel, low metallicity Mn data obtained in LTE approximation (e.g. \citealt{Eitner20}, maroon crosses; \citealt{Lai08}, grey crosses) are considered. With these data, [Mn/Fe] exhibits a $\sim$-0.6 dex plateau at low metallicity, with a steep rise after [Fe/H]$\sim$-1 dex. This trend favours the models in which the majority of SN Ia progenitors have near-$M_{ch}$. In fact, the best model from the statistical test (green line), as well as the other models with good data-model agreement (yellow lines), adopt a near-$M_{ch}$ fraction $\geq$0.75. These results are not particularly affected by the slight underestimation of [Mn/Fe] at low metallicity by the model. In fact, we test whether an increment of 25\% of adopted Mn CC-SN yields (which give a better agreement between the models and the observed low metallicity plateau) can change our conclusion, finding no substantial differences. 

The situation changes instead in the lower panel of Figure \ref{f:MnFe_combo}, where we consider both Mn data with NLTE corrections (\citealt{Eitner20}, green dots)  or LTE data from MnII lines (\citealt{Lai08}, grey dots). In fact, LTE abundances derived from MnII lines are very similar to the results obtained in NLTE (\citealt{Eitner20}).  
In order to reproduce the observed low metallicity trend, in Figure \ref{f:MnFe_combo} lower panel the Mn yields from CC-SNe are multiplied by a factor of 3: this operation is actually equivalent to adopt \citet{Koba06} yields with hypernova fraction $\epsilon_{HY}=$0 (in this work we adopt $\epsilon_{HY}=$1, as in \citealt{Romano10} best model), suggested by \citet{Eitner20} to reproduce Mn NLTE data.  To show better the trend described by \citet{Eitner20} data, we give to the \citet{Nissen00} and \citet{Adibekyan12} LTE MnI data (cyan and blue dots, respectively) a more shaded style. In fact,  we still see a discrepancy between LTE and NLTE data even at [Fe/H]$\gtrsim$-1 dex, where NLTE effects are usually lower. Unfortunately, we do not have a comparison between LTE and NLTE for [Fe/H]>-0.3 dex: thus, we cannot say if the steep rise in [Mn/Fe] up to supersolar metallicities is real or caused by observational effects (LTE-NLTE matter).\\
At variance with the upper panel, the behaviour of the observational data is followed by models with a majority of sub-$M_{ch}$ DD progenitors (50-75\%, see yellow lines in Figure \ref{f:MnFe_combo} lower panel), in agreement with recent results for the MW (\citealt{Eitner20}).
Despite of this, we can not exclude a larger contribution from SNe exploding via DDT mechanism for two reasons. First, the spread and the uncertainties\footnote{At variance with other works, \citet{Eitner20} errobars consider not only the error due to the uncertainties in stellar parameters. They sum in quadrature also the statistical error (reflecting the imperfection in the observational data). Without this latter source of error, the errorbars in Figure \ref{f:MnFe_combo} would be similar to those of \citet{Nissen00} and \citet{Adibekyan12}.} found for NLTE data does not allow to exclude a larger fraction of near-$M_{ch}$ SNe Ia. Moreover, the resulting best model for Figure \ref{f:MnFe_combo} lower panel (green line) is obtained adopting a predominant fraction (75\%) of DDT,  low WD central density (i.e. $1\times10^9$gr cm$^{-3}$) yields.
However, this is not necessarily against the conclusion made by \citet{Eitner20}. DDT \citetalias{Leung18} models with low WD central density correspond to WD with masses $<1.35$ M$_\odot$, which can be still considered as sub-Chandrasekar masses (\citealt{Leung18}). In fact, at masses $\sim 1.30$M$_\odot$ the surface He detonation may trigger a C deflagration via shock compression of the central region (\citealt{Leung18}). Moreover, as explained in Section \ref{ss:single_yields}, the abundance pattern of these low density models is much more similar to sub-$M_{ch}$ DD models than other near-$M_{ch}$ DDT models. \\
We note also that the different nucleosynthetic results by low density \citet{Seit13} models (no [Mn/Fe] decrease relative to the benchmark model) are not directly comparable with this scenario, because of the different WD mass ($>1.35$M$_\odot$) and conditions in \citet{Seit13} simulations.\\

\subsubsection*{Nickel}
Also the evolution of [Ni/Fe] vs. [Fe/H] can be used to identify the relative contribution of the different SN Ia progenitor classes (e.g. \citealt{Seit13b,Kirby19}).\\
As for Mn the lower [Ni/Fe] yields for sub-$M_{ch}$ explosions relative to near-$M_{ch}$ DDT yields are evident (see Figures \ref{f:yields}, \ref{f:NiFe}). In turn, these latter are lower than those from pure deflagration models that are indicative of SNe Iax.

From the observational point of view, Ni abundances at disposal are all inferred assuming LTE. In fact, there is little known on departures from LTE of Ni in stellar atmospheres. However, significant departures ($>0.1$dex) from LTE are not expected for Ni lines in the optical range (\citealt{Scott15,Jofre15}).

\begin{figure}
    \centering
    \includegraphics[width=1.\columnwidth]{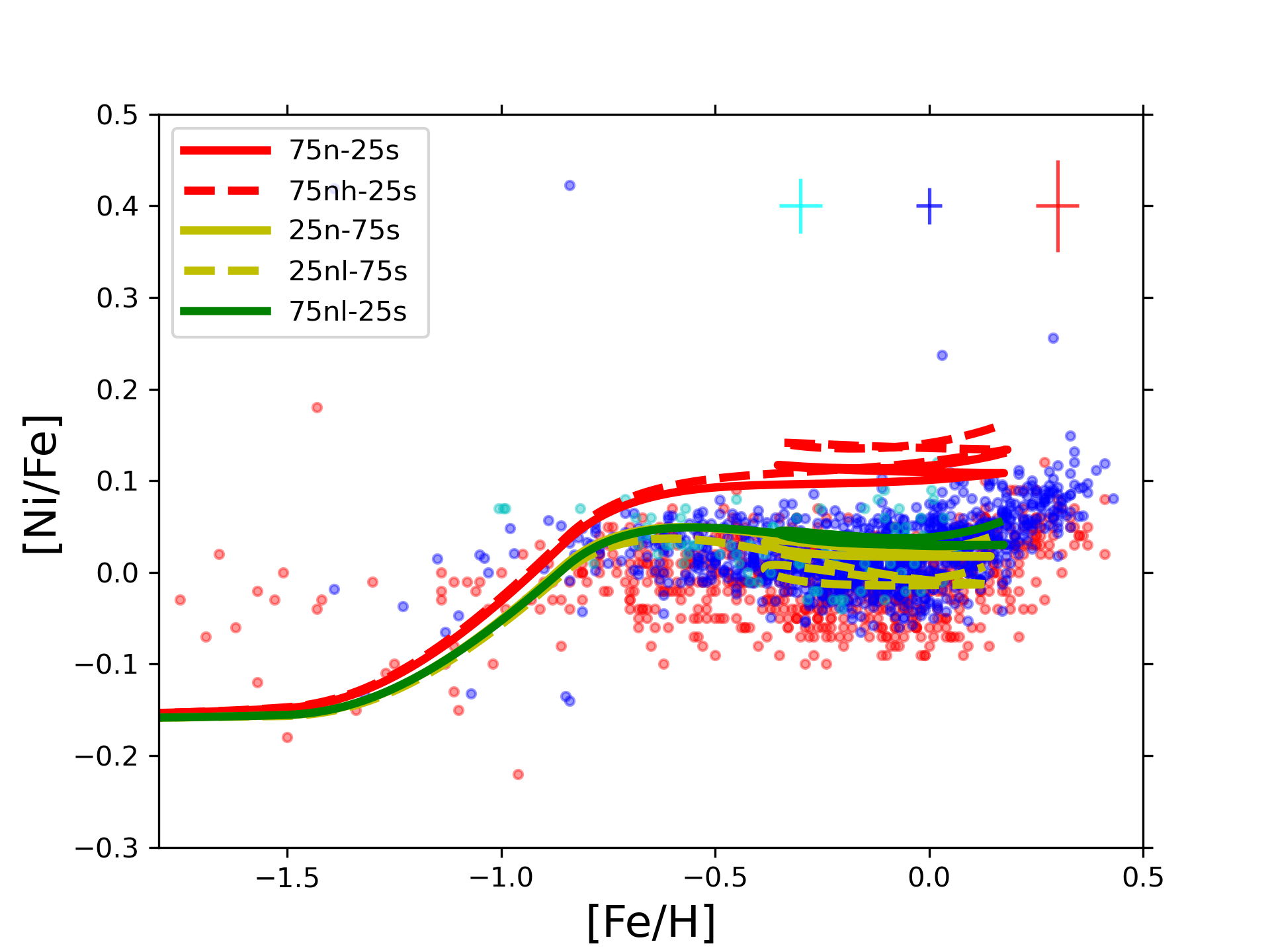}
    \caption{Same of Figure \ref{f:VFe_combo} but for [Ni/Fe]. Ni LTE data are from \citet{Chen00} (cyan points), \citet{Adibekyan12} (blue points) and \citet{Bensby14} (red points).  Cyan, blue and red errorbars indicate the typical uncertainties in \citet{Chen00}, \citet{Adibekyan12} and \citet{Bensby14} samples.}
    \label{f:NiFe_combo}
\end{figure}

Looking at Figure \ref{f:NiFe_combo}, it is suggested a large contribution ($\gtrsim$50\%) from sub-$M_{ch}$ progenitors. Models assuming a predominant contribution from benchmark near-$M_{ch}$ models (e.g. solid and dashed red lines in Figure \ref{f:NiFe_combo}) are not able to reproduce the flat [Ni/Fe] trend from the observations.\\ 
However, as for Mn the contribution to the chemical enrichment from sub-$M_{ch}$ DD progenitors can be lower if a large fraction of DDT SNe Ia has lower WD progenitor densities, as demonstrated by the best model for the abundance diagram (green line in Figure \ref{f:NiFe_combo}). This is driven by the fact that abundance patterns produced by low density DDT models are similar to those of DD models (remember Section \ref{ss:single_yields}).
Nonetheless, we noted in the previous paragraph that this does not alter our conclusion on the predominance of sub-$M_{ch}$ progenitors contribution to the chemical enrichment. \\
At variance with Cr and Mn, the behaviour of decreasing [Ni/Fe] yield with decreasing central WD density holds also for \citet{Seit13} DDT models. However, we remind that we cannot use \citet{Seit13} low density yields as an additional argument to sub-$M_{ch}$ predominance. In fact, the conditions in the simulated WD are different than in \citet{Leung18}.

\section{Summary and discussion}
\label{s:conclusions}

\begin{figure*}
    \centering
    \includegraphics[width=0.85\textwidth]{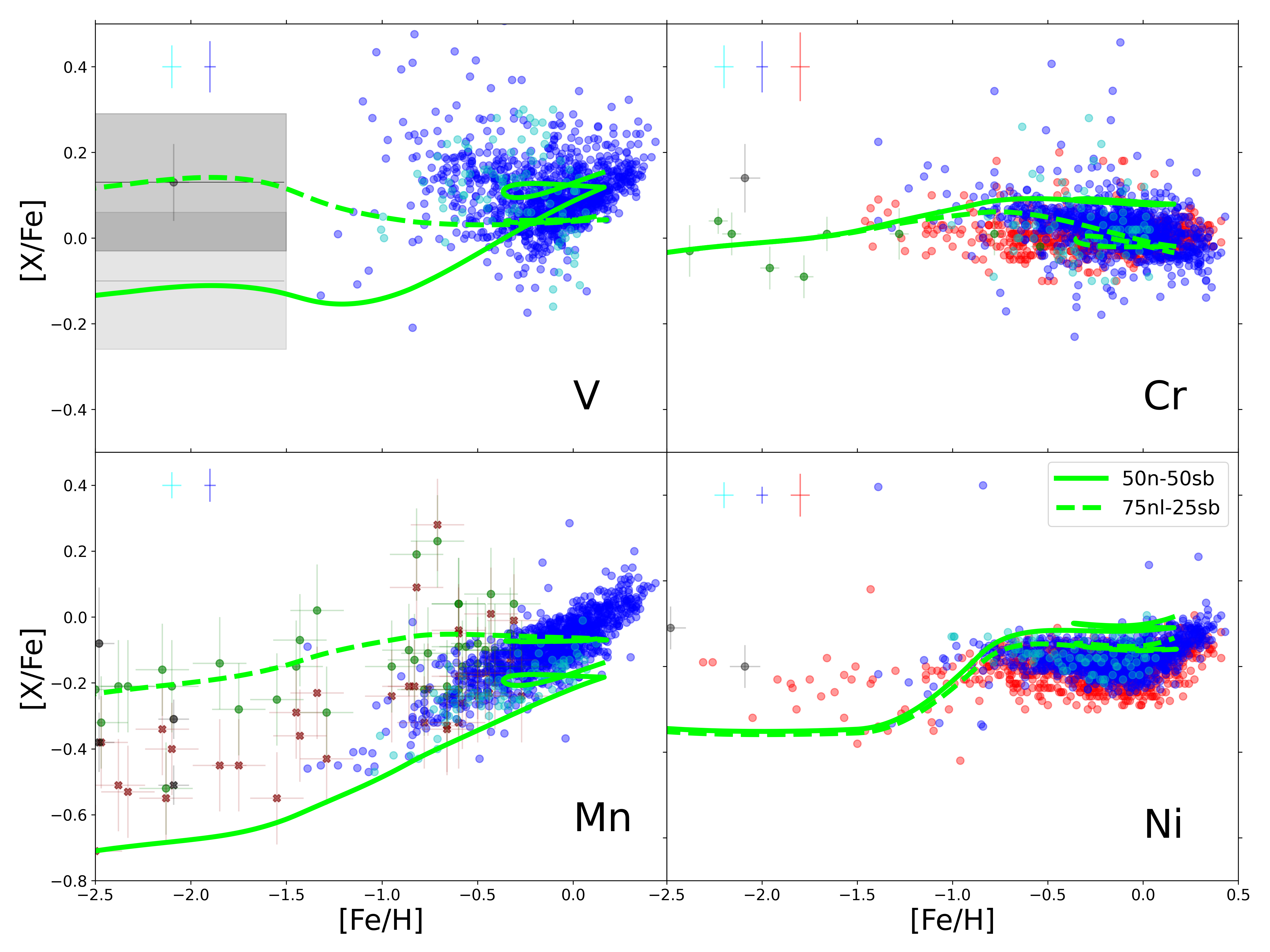}
    \caption{[X/Fe] vs. [Fe/H] ratios predicted by the overall best models obtained adopting the statistical test described in Section \ref{ss:combo_yields}. Green solid lines indicate the best model for standard \citet{Koba06} CC-SN yields, while green dashed lines stand for the best model obtained with modified CC-SN yields for V and Mn tuned to reproduce low-metallicity data. For the observational data, the colour code is the same as described in Figures \ref{f:VFe_combo}, \ref{f:CrFe_combo}, \ref{f:MnFe_combo}, \ref{f:NiFe_combo}.}
    \label{f:BEST_MODEL}
\end{figure*}

\begin{table*}
    \centering
    \caption{ Predicted $\log$(X/H)+12 solar abundances for Fe-peak elements by our two-infall best models using combinations of SN Ia yield sets. Model abundances are taken at $t=9.25$ Gyr in order to take the time at which the protosolar cloud was formed. The predictions are compared to observed photospheric solar abundances by \citet{Asplund09}.}
    \begin{tabular}{c | c  c  c  c  c  }
        \hline
        \hline
        Observation  & $\log$(Fe/H)+12 & $\log$(V/H)+12 & $\log$(Cr/H)+12 & $\log$(Mn/H)+12 & $\log$(Ni/H)+12  \\
        \hline
        \citet{Asplund09} & 7.50$\pm$0.04 & 3.93$\pm$0.08 & 5.64$\pm$0.04 & 5.43$\pm$0.04 & 6.22$\pm$0.04 \\
        \hline
        Model & $\log$(Fe/H)+12 & $\log$(V/H)+12 & $\log$(Cr/H)+12 & $\log$(Mn/H)+12 & $\log$(Ni/H)+12\\
        \hline
        50n-50sb & 7.48 & 4.03 & 5.71 & 5.24 & 6.31\\
        75nl-25sb$^{(1)}$ &	7.49 & 3.95 & 5.61 & 5.35 & 6.26 \\ 
         \hline
         \hline
    \end{tabular}\\
    $^{(1)}$ values for V and Mn are given adopting modified CC-SN yields tuned to reproduce low-metallicity data.
    \label{t:Fe_abundances_comboBEST}
\end{table*}

In this paper we have investigated the effects of a broad compilation of Type Ia SN (SN Ia) yield sets available in the literature on the MW chemical evolution. In particular, we test the near-Chandrasekar (near-$M_{ch}$) WD, delayed detonation (DDT) yields from \citet{Seit13} and \citet{Leung18}, the near-$M_{ch}$, pure deflagration (PTD) yields from \citet{Fink14} and \citet{Leung18} (which can be representative of SNe Iax) and the sub-Chandrasekar (sub-$M_{ch}$), double detonation (DD) yields from \citet{Shen18} and \citet{Leung20}. These yields are also compared with the "standard" yields adopted in chemical evolution, i.e. the W7 and WDD2 models (in the updated \citealt{Leung18} version). Moreover, we have combined the yields from different progenitor classes (i.e. near-$M_{ch}$ with sub-$M_{ch}$) in order to asses the dominant SN Ia class in terms of chemical enrichment.\\
We have assumed a specific delay-time-distribution function (DTD) for SN Ia, i.e.  the single degenerate from \citet{Matteucci01}. The results would not have changed if a double degenerate DTD (e.g. \citealt{Greggio05}) or a $t^{-\sim1}$ DTD (e.g. \citealt{Totani08}) would have been adopted, since they predict very similar SN Ia rates. This has been shown in \citet{Matteucci09}, where it was shown that the DTDs are quite similar and produce similar chemical results.\\
We have run detailed one-infall and two-infall models for MW chemical evolution, in order to highlight the differences between yields in the first case and to better reproduce the observations in the second one.

We have mostly concentrated on elements whose production by SNe Ia is important. In particular, we have looked at vanadium, chromium, manganese and nickel, where the differences among the yield sets adopted in this work are important.\\
In order to test the results obtained by means of different sets/combination of yields, we have adopted recent observational data from the literature, exploiting NLTE corrected measurements when possible.
The best models arising from the comparison theory-observations are determined also quantitatively by adopting a statistical test.\\

The main results of this work can be summarised as follows:
\begin{enumerate}
    \item Despite of being less physical than more recent multi-dimensional models, W7 and WDD2 models can still be safely adopted in chemical evolution models if the goal of the study is the total abundance of iron or the [$\alpha$/Fe] vs. [Fe/H] diagram. In fact, Fe yields (and so [$\alpha$/Fe] ratios) are not particularly different for most of multi dimensional near-$M_{ch}$ and sub-$M_{ch}$ yields tested in this work. 
    \item Only Mn and Ni show a clear distinct behaviour in the abundance ratio vs. metallicity diagram for all the three explosion mechanisms investigated in this work. In particular, pure deflagration (PTD) models produce higher [Mn,Ni/Fe] ratios relative to delayed detonation (DDT) models, which in turn produce higher ratios than sub-$M_{ch}$ double detonation (DD) models.
    \item In the case of Cr and V, the main differences can be instead caused by the different initial conditions of the exploding WD (WD central density for near-$M_{ch}$ models, He detonation pattern for sub-$M_{ch}$ models) or by the different results obtained by different SN Ia simulations sharing similar conditions (e.g. \citealt{Shen18,Leung20}). 
    \item By adopting combinations of different SN Ia progenitor classes, we note that the [Ni/Fe] abundance diagram suggests an important contribution ($\gtrsim$50\%) by sub-$M_{ch}$ SNe Ia to Galactic chemical evolution. A similar suggestion can be found for [Mn/Fe] if we consider enhanced CC-SN Mn yields, which best fit NLTE measurements at low metallicity.
    A minor contribution by double detonating sub-$M_{ch}$ WD to Ni and Mn can be achieved by adopting near-$M_{ch}$ DDT yields by \citet{Leung18} with lower WD central density.  The results in this case are almost indistinguishable from the scenario in which we have a larger fraction of sub-$M_{ch}$ WD combined with benchmark near-$M_{ch}$ WD models.\\
    In any case, the results for Mn are in fair agreement with previous work by \citet{Eitner20}. In fact, low density models by \citet{Leung18} can already be considered as sub-Chandrasekar mass models (\citealt{Leung18}). Concerning the results of \citet{Koba20}, who claimed a low-to-negligible contribution by sub-$M_{ch}$ WDs for Mn and Ni, we point out that they do not consider NLTE abundances for Mn. This can strongly affect the conclusions regarding that element, as we have seen in Figure \ref{f:MnFe_combo}. For Ni instead, \citet{Koba20} different result can be explained by the fact that they adopted a mixture of \citet{Leung20} sub-$M_{ch}$ yields of different masses with spherical He detonation, which leave lower overall [Ni/Fe] (see Figure 18 of \citealt{Koba20}).
    \item Concerning V and Cr, the comparison with observations suggests that for these two elements a low but non negligible fraction ($\sim$25\%) of SNe Ia may come from sub-$M_{ch}$ WD with an aspherical He detonation pattern (i.e. bubble detonation in the models of \citealt{Leung20}).  Even if different near-$M_{ch}$ models are able to explain the behaviour of the two elements separately, the general picture is better explained by a combination of progenitors including sub-$M_{ch}$ WD with bubble detonation.\\ 
    However, we remind that uncertainties in the abundances observed at low metallicity (for V, \citealt{Ou20}) as well as in the SN Ia yields themselves (for Cr, \citealt{Shen18,Leung20}) place limits on our analysis of SNe Ia contribution to these elements.
    \item The overall best models found by adopting the statistical test described in Section \ref{ss:combo_yields} are shown in Figure \ref{f:BEST_MODEL}, with their predicted solar abundances listed in Table \ref{t:Fe_abundances_comboBEST}.  In particular, we find that the best model obtained adopting standard CC-SN yields from the literature (\citealt{Koba06}) is an equal distribution of near-$M_{ch}$ and sub-$M_{ch}$ SNe Ia. When we adopt modified CC-SN yields for V and Mn (to fit low metallicity data), instead, we find a combination with 75\% of near-$M_{ch}$ and 25\% of sub-$M_{ch}$ models. In this case, however, the near-$M_{ch}$ fraction is constituted by low density DDT models, which have similar yields to sub-$M_{ch}$ DD models (see Section \ref{ss:single_yields}). This yield combination leads to very similar results to the case in which we have a predominant fraction of sub-$M_{ch}$ DD models combined with near-$M_{ch}$ benchmark DDT models. Concerning the sub-$M_{ch}$ fraction, both the best models suggest the adoption of models with bubble He detonation pattern, in agreement with point (v).\\
    The analysis of Figure \ref{f:BEST_MODEL} and Table \ref{t:Fe_abundances_comboBEST} also highlights that the best model with modified CC-SN yields fits quite well the observed abundance for all the elements considered at variance with the best model with standard CC-SN yields,  for which a $\lq$golden combination' is not really found. This result may perhaps indicate that NLTE and singly ionised abundances at low metallicity should be definitely preferred. In turn, this can also provide tighter constraints on CC-SN yields, in particular for elements such as V and Mn.

\end{enumerate}

Our study provides a view of the influence of different SN Ia yields on the chemical evolution of the solar neighbourhood. \\
It would be natural to extend this study to other environments, such as dwarf MW satellites (e.g. \citealt{Kirby19,DeLosReyes20,Koba20}) or the Galactic bulge. In this way, we can study in more detail the metallicity dependence of the different SN Ia yields. Moreover, these objects could be very useful to understand if different environments lead to a different contribution from SN Ia subclasses or not.

\section*{Acknowledgements}
The author thanks F. Matteucci for the fruitful suggestions and for the help in the writing of the manuscript and D. Romano, E. Spitoni, F. Vincenzo for the useful comments.  The author also thanks the anonymous referee for careful reading of the manuscript and useful suggestions.

\section*{Data availability}
The data underlying this article will be shared on reasonable request to the corresponding author.




\bibliographystyle{mnras}
\bibliography{SNIa_paper}

\begin{thebibliography}{}
\makeatletter
\relax
\def\mn@urlcharsother{\let\do\@makeother \do\$\do\&\do\#\do\^\do\_\do\%\do\~}
\def\mn@doi{\begingroup\mn@urlcharsother \@ifnextchar [ {\mn@doi@}
  {\mn@doi@[]}}
\def\mn@doi@[#1]#2{\def\@tempa{#1}\ifx\@tempa\@empty \href
  {http://dx.doi.org/#2} {doi:#2}\else \href {http://dx.doi.org/#2} {#1}\fi
  \endgroup}
\def\mn@eprint#1#2{\mn@eprint@#1:#2::\@nil}
\def\mn@eprint@arXiv#1{\href {http://arxiv.org/abs/#1} {{\tt arXiv:#1}}}
\def\mn@eprint@dblp#1{\href {http://dblp.uni-trier.de/rec/bibtex/#1.xml}
  {dblp:#1}}
\def\mn@eprint@#1:#2:#3:#4\@nil{\def\@tempa {#1}\def\@tempb {#2}\def\@tempc
  {#3}\ifx \@tempc \@empty \let \@tempc \@tempb \let \@tempb \@tempa \fi \ifx
  \@tempb \@empty \def\@tempb {arXiv}\fi \@ifundefined
  {mn@eprint@\@tempb}{\@tempb:\@tempc}{\expandafter \expandafter \csname
  mn@eprint@\@tempb\endcsname \expandafter{\@tempc}}}

\bibitem[\protect\citeauthoryear{{Adibekyan}, {Sousa}, {Santos}, {Delgado
  Mena}, {Gonz{\'a}lez Hern{\'a}ndez}, {Israelian}, {Mayor}  \&
  {Khachatryan}}{{Adibekyan} et~al.}{2012}]{Adibekyan12}
{Adibekyan} V.~Z.,  {Sousa} S.~G.,  {Santos} N.~C.,  {Delgado Mena} E.,
  {Gonz{\'a}lez Hern{\'a}ndez} J.~I.,  {Israelian} G.,  {Mayor} M.,
  {Khachatryan} G.,  2012, \mn@doi [\aap] {10.1051/0004-6361/201219401}, \href
  {https://ui.adsabs.harvard.edu/abs/2012A&A...545A..32A} {545, A32}

\bibitem[\protect\citeauthoryear{{Arnett}}{{Arnett}}{1996}]{Arnett96}
{Arnett} D.,  1996, {Supernovae and Nucleosynthesis: An Investigation of the
  History of Matter from the Big Bang to the Present}

\bibitem[\protect\citeauthoryear{{Asplund}, {Grevesse}, {Sauval}  \&
  {Scott}}{{Asplund} et~al.}{2009}]{Asplund09}
{Asplund} M.,  {Grevesse} N.,  {Sauval} A.~J.,   {Scott} P.,  2009, \mn@doi
  [\araa] {10.1146/annurev.astro.46.060407.145222}, \href
  {https://ui.adsabs.harvard.edu/abs/2009ARA&A..47..481A} {47, 481}

\bibitem[\protect\citeauthoryear{{Bensby}, {Feltzing}  \& {Oey}}{{Bensby}
  et~al.}{2014}]{Bensby14}
{Bensby} T.,  {Feltzing} S.,   {Oey} M.~S.,  2014, \mn@doi [\aap]
  {10.1051/0004-6361/201322631}, \href
  {https://ui.adsabs.harvard.edu/abs/2014A&A...562A..71B} {562, A71}

\bibitem[\protect\citeauthoryear{{Bergemann} \& {Cescutti}}{{Bergemann} \&
  {Cescutti}}{2010}]{Bergemann10Cr}
{Bergemann} M.,  {Cescutti} G.,  2010, \mn@doi [\aap]
  {10.1051/0004-6361/201014250}, \href
  {https://ui.adsabs.harvard.edu/abs/2010A&A...522A...9B} {522, A9}

\bibitem[\protect\citeauthoryear{{Bergemann} \& {Gehren}}{{Bergemann} \&
  {Gehren}}{2008}]{Bergemann08}
{Bergemann} M.,  {Gehren} T.,  2008, \mn@doi [\aap]
  {10.1051/0004-6361:200810098}, \href
  {https://ui.adsabs.harvard.edu/abs/2008A&A...492..823B} {492, 823}

\bibitem[\protect\citeauthoryear{{Bergemann} et~al.,}{{Bergemann}
  et~al.}{2019}]{Bergemann19}
{Bergemann} M.,  et~al., 2019, \mn@doi [\aap] {10.1051/0004-6361/201935811},
  \href {https://ui.adsabs.harvard.edu/abs/2019A&A...631A..80B} {631, A80}

\bibitem[\protect\citeauthoryear{{Bonifacio} et~al.,}{{Bonifacio}
  et~al.}{2009}]{Bonifacio09}
{Bonifacio} P.,  et~al., 2009, \mn@doi [\aap] {10.1051/0004-6361/200810610},
  \href {https://ui.adsabs.harvard.edu/abs/2009A&A...501..519B} {501, 519}

\bibitem[\protect\citeauthoryear{{Cayrel} et~al.,}{{Cayrel}
  et~al.}{2004}]{Cayrel04}
{Cayrel} R.,  et~al., 2004, \mn@doi [\aap] {10.1051/0004-6361:20034074}, \href
  {https://ui.adsabs.harvard.edu/abs/2004A&A...416.1117C} {416, 1117}

\bibitem[\protect\citeauthoryear{{Cescutti}}{{Cescutti}}{2008}]{Cescutti08}
{Cescutti} G.,  2008, \mn@doi [\aap] {10.1051/0004-6361:20078571}, \href
  {https://ui.adsabs.harvard.edu/abs/2008A&A...481..691C} {481, 691}

\bibitem[\protect\citeauthoryear{{Cescutti} \& {Kobayashi}}{{Cescutti} \&
  {Kobayashi}}{2017}]{Cescutti17}
{Cescutti} G.,  {Kobayashi} C.,  2017, \mn@doi [\aap]
  {10.1051/0004-6361/201731398}, \href
  {https://ui.adsabs.harvard.edu/abs/2017A&A...607A..23C} {607, A23}

\bibitem[\protect\citeauthoryear{{Chen}, {Nissen}, {Zhao}, {Zhang}  \&
  {Benoni}}{{Chen} et~al.}{2000}]{Chen00}
{Chen} Y.~Q.,  {Nissen} P.~E.,  {Zhao} G.,  {Zhang} H.~W.,   {Benoni} T.,
  2000, \mn@doi [\aaps] {10.1051/aas:2000124}, \href
  {https://ui.adsabs.harvard.edu/abs/2000A&AS..141..491C} {141, 491}

\bibitem[\protect\citeauthoryear{{Chiappini}, {Matteucci}  \&
  {Gratton}}{{Chiappini} et~al.}{1997}]{Chiappini97}
{Chiappini} C.,  {Matteucci} F.,   {Gratton} R.,  1997, \mn@doi [\apj]
  {10.1086/303726}, \href
  {https://ui.adsabs.harvard.edu/abs/1997ApJ...477..765C} {477, 765}

\bibitem[\protect\citeauthoryear{{Eitner}, {Bergemann}, {Hansen}, {Cescutti},
  {Seitenzahl}, {Larsen}  \& {Plez}}{{Eitner} et~al.}{2020}]{Eitner20}
{Eitner} P.,  {Bergemann} M.,  {Hansen} C.~J.,  {Cescutti} G.,  {Seitenzahl}
  I.~R.,  {Larsen} S.,   {Plez} B.,  2020, \mn@doi [\aap]
  {10.1051/0004-6361/201936603}, \href
  {https://ui.adsabs.harvard.edu/abs/2020A&A...635A..38E} {635, A38}

\bibitem[\protect\citeauthoryear{{Ernandes}, {Barbuy}, {Fria{\c{c}}a}, {Hill},
  {Zoccali}, {Minniti}, {Renzini}  \& {Ortolani}}{{Ernandes}
  et~al.}{2020}]{Ernandes20}
{Ernandes} H.,  {Barbuy} B.,  {Fria{\c{c}}a} A.,  {Hill} V.,  {Zoccali} M.,
  {Minniti} D.,  {Renzini} A.,   {Ortolani} S.,  2020, arXiv e-prints, \href
  {https://ui.adsabs.harvard.edu/abs/2020arXiv200700397E} {p. arXiv:2007.00397}

\bibitem[\protect\citeauthoryear{{Fink} et~al.,}{{Fink} et~al.}{2014}]{Fink14}
{Fink} M.,  et~al., 2014, \mn@doi [\mnras] {10.1093/mnras/stt2315}, \href
  {https://ui.adsabs.harvard.edu/abs/2014MNRAS.438.1762F} {438, 1762}

\bibitem[\protect\citeauthoryear{{Gratton}, {Carretta}, {Claudi}, {Lucatello}
  \& {Barbieri}}{{Gratton} et~al.}{2003}]{Gratton03}
{Gratton} R.~G.,  {Carretta} E.,  {Claudi} R.,  {Lucatello} S.,   {Barbieri}
  M.,  2003, \mn@doi [\aap] {10.1051/0004-6361:20030439}, \href
  {https://ui.adsabs.harvard.edu/abs/2003A&A...404..187G} {404, 187}

\bibitem[\protect\citeauthoryear{{Greggio}}{{Greggio}}{2005}]{Greggio05}
{Greggio} L.,  2005, \mn@doi [\aap] {10.1051/0004-6361:20052926}, \href
  {https://ui.adsabs.harvard.edu/abs/2005A&A...441.1055G} {441, 1055}

\bibitem[\protect\citeauthoryear{{Grisoni}, {Spitoni}  \&
  {Matteucci}}{{Grisoni} et~al.}{2018}]{Grisoni18}
{Grisoni} V.,  {Spitoni} E.,   {Matteucci} F.,  2018, \mn@doi [\mnras]
  {10.1093/mnras/sty2444}, \href
  {https://ui.adsabs.harvard.edu/abs/2018MNRAS.481.2570G} {481, 2570}

\bibitem[\protect\citeauthoryear{{Hayden} et~al.,}{{Hayden}
  et~al.}{2015}]{Hayden15}
{Hayden} M.~R.,  et~al., 2015, \mn@doi [\apj] {10.1088/0004-637X/808/2/132},
  \href {https://ui.adsabs.harvard.edu/abs/2015ApJ...808..132H} {808, 132}

\bibitem[\protect\citeauthoryear{{Helmi}}{{Helmi}}{2020}]{Helmi20}
{Helmi} A.,  2020, \mn@doi [\araa] {10.1146/annurev-astro-032620-021917}, \href
  {https://ui.adsabs.harvard.edu/abs/2020ARA&A..58..205H} {58, 205}

\bibitem[\protect\citeauthoryear{{Hillebrandt} \& {Niemeyer}}{{Hillebrandt} \&
  {Niemeyer}}{2000}]{Hillebrandt00}
{Hillebrandt} W.,  {Niemeyer} J.~C.,  2000, \mn@doi [\araa]
  {10.1146/annurev.astro.38.1.191}, \href
  {https://ui.adsabs.harvard.edu/abs/2000ARA&A..38..191H} {38, 191}

\bibitem[\protect\citeauthoryear{{Hillebrandt}, {Kromer}, {R{\"o}pke}  \&
  {Ruiter}}{{Hillebrandt} et~al.}{2013}]{Hillebrandt13}
{Hillebrandt} W.,  {Kromer} M.,  {R{\"o}pke} F.~K.,   {Ruiter} A.~J.,  2013,
  \mn@doi [Frontiers of Physics] {10.1007/s11467-013-0303-2}, \href
  {https://ui.adsabs.harvard.edu/abs/2013FrPhy...8..116H} {8, 116}

\bibitem[\protect\citeauthoryear{{Hoeflich} \& {Khokhlov}}{{Hoeflich} \&
  {Khokhlov}}{1996}]{Hoeflich96}
{Hoeflich} P.,  {Khokhlov} A.,  1996, \mn@doi [\apj] {10.1086/176748}, \href
  {https://ui.adsabs.harvard.edu/abs/1996ApJ...457..500H} {457, 500}

\bibitem[\protect\citeauthoryear{{Iben} \& {Tutukov}}{{Iben} \&
  {Tutukov}}{1984}]{Iben84}
{Iben} I. J.,  {Tutukov} A.~V.,  1984, \mn@doi [\apjs] {10.1086/190932}, \href
  {https://ui.adsabs.harvard.edu/abs/1984ApJS...54..335I} {54, 335}

\bibitem[\protect\citeauthoryear{{Iben} \& {Tutukov}}{{Iben} \&
  {Tutukov}}{1991}]{Iben91}
{Iben} Icko J.,  {Tutukov} A.~V.,  1991, \mn@doi [\apj] {10.1086/169848}, \href
  {https://ui.adsabs.harvard.edu/abs/1991ApJ...370..615I} {370, 615}

\bibitem[\protect\citeauthoryear{{Iwamoto}, {Brachwitz}, {Nomoto}, {Kishimoto},
  {Umeda}, {Hix}  \& {Thielemann}}{{Iwamoto} et~al.}{1999}]{Iwa99}
{Iwamoto} K.,  {Brachwitz} F.,  {Nomoto} K.,  {Kishimoto} N.,  {Umeda} H.,
  {Hix} W.~R.,   {Thielemann} F.-K.,  1999, \mn@doi [\apjs] {10.1086/313278},
  \href {https://ui.adsabs.harvard.edu/abs/1999ApJS..125..439I} {125, 439}

\bibitem[\protect\citeauthoryear{{Jofr{\'e}} et~al.,}{{Jofr{\'e}}
  et~al.}{2015}]{Jofre15}
{Jofr{\'e}} P.,  et~al., 2015, \mn@doi [\aap] {10.1051/0004-6361/201526604},
  \href {https://ui.adsabs.harvard.edu/abs/2015A&A...582A..81J} {582, A81}

\bibitem[\protect\citeauthoryear{{Karakas}}{{Karakas}}{2010}]{Karakas10}
{Karakas} A.~I.,  2010, \mn@doi [\mnras] {10.1111/j.1365-2966.2009.16198.x},
  \href {https://ui.adsabs.harvard.edu/abs/2010MNRAS.403.1413K} {403, 1413}

\bibitem[\protect\citeauthoryear{{Kennicutt}}{{Kennicutt}}{1998}]{Kennicutt98}
{Kennicutt} Robert~C. J.,  1998, \mn@doi [\apj] {10.1086/305588}, \href
  {https://ui.adsabs.harvard.edu/abs/1998ApJ...498..541K} {498, 541}

\bibitem[\protect\citeauthoryear{{Khokhlov}}{{Khokhlov}}{1991}]{Khokhlov91}
{Khokhlov} A.~M.,  1991, \aap, \href
  {https://ui.adsabs.harvard.edu/abs/1991A&A...245..114K} {245, 114}

\bibitem[\protect\citeauthoryear{{Kirby} et~al.,}{{Kirby}
  et~al.}{2019}]{Kirby19}
{Kirby} E.~N.,  et~al., 2019, \mn@doi [\apj] {10.3847/1538-4357/ab2c02}, \href
  {https://ui.adsabs.harvard.edu/abs/2019ApJ...881...45K} {881, 45}

\bibitem[\protect\citeauthoryear{{Kobayashi}, {Umeda}, {Nomoto}, {Tominaga}  \&
  {Ohkubo}}{{Kobayashi} et~al.}{2006}]{Koba06}
{Kobayashi} C.,  {Umeda} H.,  {Nomoto} K.,  {Tominaga} N.,   {Ohkubo} T.,
  2006, \mn@doi [\apj] {10.1086/508914}, \href
  {https://ui.adsabs.harvard.edu/abs/2006ApJ...653.1145K} {653, 1145}

\bibitem[\protect\citeauthoryear{{Kobayashi}, {Nomoto}  \&
  {Hachisu}}{{Kobayashi} et~al.}{2015}]{Koba15}
{Kobayashi} C.,  {Nomoto} K.,   {Hachisu} I.,  2015, \mn@doi [\apjl]
  {10.1088/2041-8205/804/1/L24}, \href
  {https://ui.adsabs.harvard.edu/abs/2015ApJ...804L..24K} {804, L24}

\bibitem[\protect\citeauthoryear{{Kobayashi}, {Leung}  \& {Nomoto}}{{Kobayashi}
  et~al.}{2020a}]{Koba20}
{Kobayashi} C.,  {Leung} S.-C.,   {Nomoto} K.,  2020a, \mn@doi [\apj]
  {10.3847/1538-4357/ab8e44}, \href
  {https://ui.adsabs.harvard.edu/abs/2020ApJ...895..138K} {895, 138}

\bibitem[\protect\citeauthoryear{{Kobayashi}, {Karakas}  \&
  {Lugaro}}{{Kobayashi} et~al.}{2020b}]{Koba20b}
{Kobayashi} C.,  {Karakas} A.~I.,   {Lugaro} M.,  2020b, \mn@doi [\apj]
  {10.3847/1538-4357/abae65}, \href
  {https://ui.adsabs.harvard.edu/abs/2020ApJ...900..179K} {900, 179}

\bibitem[\protect\citeauthoryear{{Kromer} et~al.,}{{Kromer}
  et~al.}{2015}]{Kromer15}
{Kromer} M.,  et~al., 2015, \mn@doi [\mnras] {10.1093/mnras/stv886}, \href
  {https://ui.adsabs.harvard.edu/abs/2015MNRAS.450.3045K} {450, 3045}

\bibitem[\protect\citeauthoryear{{Kroupa}, {Tout}  \& {Gilmore}}{{Kroupa}
  et~al.}{1993}]{Kroupa93}
{Kroupa} P.,  {Tout} C.~A.,   {Gilmore} G.,  1993, \mn@doi [\mnras]
  {10.1093/mnras/262.3.545}, \href
  {https://ui.adsabs.harvard.edu/abs/1993MNRAS.262..545K} {262, 545}

\bibitem[\protect\citeauthoryear{{Lai}, {Bolte}, {Johnson}, {Lucatello},
  {Heger}  \& {Woosley}}{{Lai} et~al.}{2008}]{Lai08}
{Lai} D.~K.,  {Bolte} M.,  {Johnson} J.~A.,  {Lucatello} S.,  {Heger} A.,
  {Woosley} S.~E.,  2008, \mn@doi [\apj] {10.1086/588811}, \href
  {https://ui.adsabs.harvard.edu/abs/2008ApJ...681.1524L} {681, 1524}

\bibitem[\protect\citeauthoryear{{Leung} \& {Nomoto}}{{Leung} \&
  {Nomoto}}{2018}]{Leung18}
{Leung} S.-C.,  {Nomoto} K.,  2018, \mn@doi [\apj] {10.3847/1538-4357/aac2df},
  \href {https://ui.adsabs.harvard.edu/abs/2018ApJ...861..143L} {861, 143}

\bibitem[\protect\citeauthoryear{{Leung} \& {Nomoto}}{{Leung} \&
  {Nomoto}}{2020a}]{Leung20}
{Leung} S.-C.,  {Nomoto} K.,  2020a, \mn@doi [\apj] {10.3847/1538-4357/ab5c1f},
  \href {https://ui.adsabs.harvard.edu/abs/2020ApJ...888...80L} {888, 80}

\bibitem[\protect\citeauthoryear{{Leung} \& {Nomoto}}{{Leung} \&
  {Nomoto}}{2020b}]{Leung20b}
{Leung} S.-C.,  {Nomoto} K.,  2020b, \mn@doi [\apj] {10.3847/1538-4357/aba1e3},
  \href {https://ui.adsabs.harvard.edu/abs/2020ApJ...900...54L} {900, 54}

\bibitem[\protect\citeauthoryear{{Mannucci}, {Della Valle}  \&
  {Panagia}}{{Mannucci} et~al.}{2006}]{Mannucci06}
{Mannucci} F.,  {Della Valle} M.,   {Panagia} N.,  2006, \mn@doi [\mnras]
  {10.1111/j.1365-2966.2006.10501.x}, \href
  {https://ui.adsabs.harvard.edu/abs/2006MNRAS.370..773M} {370, 773}

\bibitem[\protect\citeauthoryear{{Maoz} \& {Graur}}{{Maoz} \&
  {Graur}}{2017}]{Maoz17}
{Maoz} D.,  {Graur} O.,  2017, \mn@doi [\apj] {10.3847/1538-4357/aa8b6e}, \href
  {https://ui.adsabs.harvard.edu/abs/2017ApJ...848...25M} {848, 25}

\bibitem[\protect\citeauthoryear{{Maoz}, {Mannucci}  \& {Nelemans}}{{Maoz}
  et~al.}{2014}]{Maoz14}
{Maoz} D.,  {Mannucci} F.,   {Nelemans} G.,  2014, \mn@doi [\araa]
  {10.1146/annurev-astro-082812-141031}, \href
  {https://ui.adsabs.harvard.edu/abs/2014ARA&A..52..107M} {52, 107}

\bibitem[\protect\citeauthoryear{{Matteucci}}{{Matteucci}}{2003}]{Matteucci03}
{Matteucci} F.,  2003, {The Chemical Evolution of the Galaxy},
  \mn@doi{10.1007/978-94-010-0967-6.
}

\bibitem[\protect\citeauthoryear{{Matteucci}}{{Matteucci}}{2012}]{Matteucci12}
{Matteucci} F.,  2012, {Chemical Evolution of Galaxies},
  \mn@doi{10.1007/978-3-642-22491-1.
}

\bibitem[\protect\citeauthoryear{{Matteucci} \& {Francois}}{{Matteucci} \&
  {Francois}}{1989}]{Matteucci89}
{Matteucci} F.,  {Francois} P.,  1989, \mn@doi [\mnras]
  {10.1093/mnras/239.3.885}, \href
  {https://ui.adsabs.harvard.edu/abs/1989MNRAS.239..885M} {239, 885}

\bibitem[\protect\citeauthoryear{{Matteucci} \& {Greggio}}{{Matteucci} \&
  {Greggio}}{1986}]{Matteucci86}
{Matteucci} F.,  {Greggio} L.,  1986, \aap, \href
  {https://ui.adsabs.harvard.edu/abs/1986A&A...154..279M} {154, 279}

\bibitem[\protect\citeauthoryear{{Matteucci} \& {Recchi}}{{Matteucci} \&
  {Recchi}}{2001}]{Matteucci01}
{Matteucci} F.,  {Recchi} S.,  2001, \mn@doi [\apj] {10.1086/322472}, \href
  {https://ui.adsabs.harvard.edu/abs/2001ApJ...558..351M} {558, 351}

\bibitem[\protect\citeauthoryear{{Matteucci} \& {Tornambe}}{{Matteucci} \&
  {Tornambe}}{1985}]{Matteucci85}
{Matteucci} F.,  {Tornambe} A.,  1985, \aap, \href
  {https://ui.adsabs.harvard.edu/abs/1985A&A...142...13M} {142, 13}

\bibitem[\protect\citeauthoryear{{Matteucci}, {Spitoni}, {Recchi}  \&
  {Valiante}}{{Matteucci} et~al.}{2009}]{Matteucci09}
{Matteucci} F.,  {Spitoni} E.,  {Recchi} S.,   {Valiante} R.,  2009, \mn@doi
  [\aap] {10.1051/0004-6361/200911869}, \href
  {https://ui.adsabs.harvard.edu/abs/2009A&A...501..531M} {501, 531}

\bibitem[\protect\citeauthoryear{{Melioli}, {Brighenti}, {D'Ercole}  \& {de
  Gouveia Dal Pino}}{{Melioli} et~al.}{2009}]{Melioli09}
{Melioli} C.,  {Brighenti} F.,  {D'Ercole} A.,   {de Gouveia Dal Pino} E.~M.,
  2009, \mn@doi [\mnras] {10.1111/j.1365-2966.2009.14725.x}, \href
  {https://ui.adsabs.harvard.edu/abs/2009MNRAS.399.1089M} {399, 1089}

\bibitem[\protect\citeauthoryear{{Niemeyer}, {Hillebrandt}  \&
  {Woosley}}{{Niemeyer} et~al.}{1996}]{Niemeyer96}
{Niemeyer} J.~C.,  {Hillebrandt} W.,   {Woosley} S.~E.,  1996, \mn@doi [\apj]
  {10.1086/178017}, \href
  {https://ui.adsabs.harvard.edu/abs/1996ApJ...471..903N} {471, 903}

\bibitem[\protect\citeauthoryear{{Nissen}, {Chen}, {Schuster}  \&
  {Zhao}}{{Nissen} et~al.}{2000}]{Nissen00}
{Nissen} P.~E.,  {Chen} Y.~Q.,  {Schuster} W.~J.,   {Zhao} G.,  2000, \aap,
  \href {https://ui.adsabs.harvard.edu/abs/2000A&A...353..722N} {353, 722}

\bibitem[\protect\citeauthoryear{{Nomoto}}{{Nomoto}}{1982}]{Nomoto82b}
{Nomoto} K.,  1982, \mn@doi [\apj] {10.1086/160031}, \href
  {https://ui.adsabs.harvard.edu/abs/1982ApJ...257..780N} {257, 780}

\bibitem[\protect\citeauthoryear{{Nomoto}, {Thielemann}  \& {Yokoi}}{{Nomoto}
  et~al.}{1984}]{Nomoto84}
{Nomoto} K.,  {Thielemann} F.~K.,   {Yokoi} K.,  1984, \mn@doi [\apj]
  {10.1086/162639}, \href
  {https://ui.adsabs.harvard.edu/abs/1984ApJ...286..644N} {286, 644}

\bibitem[\protect\citeauthoryear{{Nomoto}, {Yamaoka}, {Shigeyama}, {Kumagai}
  \& {Tsujimoto}}{{Nomoto} et~al.}{1994}]{Nomoto94}
{Nomoto} K.,  {Yamaoka} H.,  {Shigeyama} T.,  {Kumagai} S.,   {Tsujimoto} T.,
  1994, in {Bludman} S.~A.,  {Mochkovitch} R.,   {Zinn-Justin} J.,  eds,
  Supernovae. p.~199

\bibitem[\protect\citeauthoryear{{Ou}, {Roederer}, {Sneden}, {Cowan}, {Lawler},
  {Shectman}  \& {Thompson}}{{Ou} et~al.}{2020}]{Ou20}
{Ou} X.,  {Roederer} I.~U.,  {Sneden} C.,  {Cowan} J.~J.,  {Lawler} J.~E.,
  {Shectman} S.~A.,   {Thompson} I.~B.,  2020, \mn@doi [\apj]
  {10.3847/1538-4357/abaa50}, \href
  {https://ui.adsabs.harvard.edu/abs/2020ApJ...900..106O} {900, 106}

\bibitem[\protect\citeauthoryear{{Pakmor}, {Kromer}, {Taubenberger}, {Sim},
  {R{\"o}pke}  \& {Hillebrandt}}{{Pakmor} et~al.}{2012}]{Pakmor12}
{Pakmor} R.,  {Kromer} M.,  {Taubenberger} S.,  {Sim} S.~A.,  {R{\"o}pke}
  F.~K.,   {Hillebrandt} W.,  2012, \mn@doi [\apjl]
  {10.1088/2041-8205/747/1/L10}, \href
  {https://ui.adsabs.harvard.edu/abs/2012ApJ...747L..10P} {747, L10}

\bibitem[\protect\citeauthoryear{{Palla}, {Matteucci}, {Spitoni}, {Vincenzo}
  \& {Grisoni}}{{Palla} et~al.}{2020a}]{Palla20b}
{Palla} M.,  {Matteucci} F.,  {Spitoni} E.,  {Vincenzo} F.,   {Grisoni} V.,
  2020a, \mn@doi [\mnras] {10.1093/mnras/staa2437}, \href
  {https://ui.adsabs.harvard.edu/abs/2020MNRAS.498.1710P} {498, 1710}

\bibitem[\protect\citeauthoryear{{Palla}, {Matteucci}, {Calura}  \&
  {Longo}}{{Palla} et~al.}{2020b}]{Palla20}
{Palla} M.,  {Matteucci} F.,  {Calura} F.,   {Longo} F.,  2020b, \mn@doi [\apj]
  {10.3847/1538-4357/ab6080}, \href
  {https://ui.adsabs.harvard.edu/abs/2020ApJ...889....4P} {889, 4}

\bibitem[\protect\citeauthoryear{{Perlmutter} et~al.,}{{Perlmutter}
  et~al.}{1999}]{Perlmutter99}
{Perlmutter} S.,  et~al., 1999, \mn@doi [\apj] {10.1086/307221}, \href
  {https://ui.adsabs.harvard.edu/abs/1999ApJ...517..565P} {517, 565}

\bibitem[\protect\citeauthoryear{{Phillips}}{{Phillips}}{1993}]{Phillips93}
{Phillips} M.~M.,  1993, \mn@doi [\apjl] {10.1086/186970}, \href
  {https://ui.adsabs.harvard.edu/abs/1993ApJ...413L.105P} {413, L105}

\bibitem[\protect\citeauthoryear{{Prantzos}, {Abia}, {Limongi}, {Chieffi}  \&
  {Cristallo}}{{Prantzos} et~al.}{2018}]{Prantzos18}
{Prantzos} N.,  {Abia} C.,  {Limongi} M.,  {Chieffi} A.,   {Cristallo} S.,
  2018, \mn@doi [\mnras] {10.1093/mnras/sty316}, \href
  {https://ui.adsabs.harvard.edu/abs/2018MNRAS.476.3432P} {476, 3432}

\bibitem[\protect\citeauthoryear{{Prochaska}, {Naumov}, {Carney}, {McWilliam}
  \& {Wolfe}}{{Prochaska} et~al.}{2000}]{Prochaska00}
{Prochaska} J.~X.,  {Naumov} S.~O.,  {Carney} B.~W.,  {McWilliam} A.,   {Wolfe}
  A.~M.,  2000, \mn@doi [\aj] {10.1086/316818}, \href
  {https://ui.adsabs.harvard.edu/abs/2000AJ....120.2513P} {120, 2513}

\bibitem[\protect\citeauthoryear{{Riess} et~al.,}{{Riess}
  et~al.}{1998}]{Riess98}
{Riess} A.~G.,  et~al., 1998, \mn@doi [\aj] {10.1086/300499}, \href
  {https://ui.adsabs.harvard.edu/abs/1998AJ....116.1009R} {116, 1009}

\bibitem[\protect\citeauthoryear{{Romano}, {Karakas}, {Tosi}  \&
  {Matteucci}}{{Romano} et~al.}{2010}]{Romano10}
{Romano} D.,  {Karakas} A.~I.,  {Tosi} M.,   {Matteucci} F.,  2010, \mn@doi
  [\aap] {10.1051/0004-6361/201014483}, \href
  {https://ui.adsabs.harvard.edu/abs/2010A&A...522A..32R} {522, A32}

\bibitem[\protect\citeauthoryear{{Ruiter}}{{Ruiter}}{2020}]{Ruiter20}
{Ruiter} A.~J.,  2020, arXiv e-prints, \href
  {https://ui.adsabs.harvard.edu/abs/2020arXiv200102947R} {p. arXiv:2001.02947}

\bibitem[\protect\citeauthoryear{{Scott}, {Asplund}, {Grevesse}, {Bergemann}
  \& {Sauval}}{{Scott} et~al.}{2015}]{Scott15}
{Scott} P.,  {Asplund} M.,  {Grevesse} N.,  {Bergemann} M.,   {Sauval} A.~J.,
  2015, \mn@doi [\aap] {10.1051/0004-6361/201424110}, \href
  {https://ui.adsabs.harvard.edu/abs/2015A&A...573A..26S} {573, A26}

\bibitem[\protect\citeauthoryear{{Seitenzahl} \& {Townsley}}{{Seitenzahl} \&
  {Townsley}}{2017}]{Seit17}
{Seitenzahl} I.~R.,  {Townsley} D.~M.,  2017, {Nucleosynthesis in Thermonuclear
  Supernovae}.
p.~1955, \mn@doi{10.1007/978-3-319-21846-5_87}

\bibitem[\protect\citeauthoryear{{Seitenzahl} et~al.,}{{Seitenzahl}
  et~al.}{2013a}]{Seit13}
{Seitenzahl} I.~R.,  et~al., 2013a, \mn@doi [\mnras] {10.1093/mnras/sts402},
  \href {https://ui.adsabs.harvard.edu/abs/2013MNRAS.429.1156S} {429, 1156}

\bibitem[\protect\citeauthoryear{{Seitenzahl}, {Cescutti}, {R{\"o}pke},
  {Ruiter}  \& {Pakmor}}{{Seitenzahl} et~al.}{2013b}]{Seit13b}
{Seitenzahl} I.~R.,  {Cescutti} G.,  {R{\"o}pke} F.~K.,  {Ruiter} A.~J.,
  {Pakmor} R.,  2013b, \mn@doi [\aap] {10.1051/0004-6361/201322599}, \href
  {https://ui.adsabs.harvard.edu/abs/2013A&A...559L...5S} {559, L5}

\bibitem[\protect\citeauthoryear{{Shen}, {Kasen}, {Miles}  \&
  {Townsley}}{{Shen} et~al.}{2018}]{Shen18}
{Shen} K.~J.,  {Kasen} D.,  {Miles} B.~J.,   {Townsley} D.~M.,  2018, \mn@doi
  [\apj] {10.3847/1538-4357/aaa8de}, \href
  {https://ui.adsabs.harvard.edu/abs/2018ApJ...854...52S} {854, 52}

\bibitem[\protect\citeauthoryear{{Silva Aguirre} et~al.,}{{Silva Aguirre}
  et~al.}{2018}]{Silva18}
{Silva Aguirre} V.,  et~al., 2018, \mn@doi [\mnras] {10.1093/mnras/sty150},
  \href {https://ui.adsabs.harvard.edu/abs/2018MNRAS.475.5487S} {475, 5487}

\bibitem[\protect\citeauthoryear{{Spitoni}, {Matteucci}, {Recchi}, {Cescutti}
  \& {Pipino}}{{Spitoni} et~al.}{2009}]{Spitoni09}
{Spitoni} E.,  {Matteucci} F.,  {Recchi} S.,  {Cescutti} G.,   {Pipino} A.,
  2009, \mn@doi [\aap] {10.1051/0004-6361/200911768}, \href
  {https://ui.adsabs.harvard.edu/abs/2009A&A...504...87S} {504, 87}

\bibitem[\protect\citeauthoryear{{Spitoni}, {Silva Aguirre}, {Matteucci},
  {Calura}  \& {Grisoni}}{{Spitoni} et~al.}{2019}]{Spitoni19}
{Spitoni} E.,  {Silva Aguirre} V.,  {Matteucci} F.,  {Calura} F.,   {Grisoni}
  V.,  2019, \mn@doi [\aap] {10.1051/0004-6361/201834188}, \href
  {https://ui.adsabs.harvard.edu/abs/2019A&A...623A..60S} {623, A60}

\bibitem[\protect\citeauthoryear{{Spitoni}, {Verma}, {Silva Aguirre}  \&
  {Calura}}{{Spitoni} et~al.}{2020}]{Spitoni20}
{Spitoni} E.,  {Verma} K.,  {Silva Aguirre} V.,   {Calura} F.,  2020, \mn@doi
  [\aap] {10.1051/0004-6361/201937275}, \href
  {https://ui.adsabs.harvard.edu/abs/2020A&A...635A..58S} {635, A58}

\bibitem[\protect\citeauthoryear{{Spitoni} et~al.,}{{Spitoni}
  et~al.}{2021}]{Spitoni21}
{Spitoni} E.,  et~al., 2021, arXiv e-prints, \href
  {https://ui.adsabs.harvard.edu/abs/2021arXiv210108803S} {p. arXiv:2101.08803}

\bibitem[\protect\citeauthoryear{{Thielemann}, {Nomoto}  \&
  {Yokoi}}{{Thielemann} et~al.}{1986}]{Thielemann86}
{Thielemann} F.~K.,  {Nomoto} K.,   {Yokoi} K.,  1986, \aap, \href
  {https://ui.adsabs.harvard.edu/abs/1986A&A...158...17T} {158, 17}

\bibitem[\protect\citeauthoryear{{Totani}, {Morokuma}, {Oda}, {Doi}  \&
  {Yasuda}}{{Totani} et~al.}{2008}]{Totani08}
{Totani} T.,  {Morokuma} T.,  {Oda} T.,  {Doi} M.,   {Yasuda} N.,  2008,
  \mn@doi [\pasj] {10.1093/pasj/60.6.1327}, \href
  {https://ui.adsabs.harvard.edu/abs/2008PASJ...60.1327T} {60, 1327}

\bibitem[\protect\citeauthoryear{{Whelan} \& {Iben}}{{Whelan} \&
  {Iben}}{1973}]{Whelan73}
{Whelan} J.,  {Iben} Icko J.,  1973, \mn@doi [\apj] {10.1086/152565}, \href
  {https://ui.adsabs.harvard.edu/abs/1973ApJ...186.1007W} {186, 1007}

\bibitem[\protect\citeauthoryear{{Woosley} \& {Weaver}}{{Woosley} \&
  {Weaver}}{1995}]{WW95}
{Woosley} S.~E.,  {Weaver} T.~A.,  1995, \mn@doi [\apjs] {10.1086/192237},
  \href {https://ui.adsabs.harvard.edu/abs/1995ApJS..101..181W} {101, 181}

\bibitem[\protect\citeauthoryear{{Yong} et~al.,}{{Yong} et~al.}{2013}]{Yong13}
{Yong} D.,  et~al., 2013, \mn@doi [\apj] {10.1088/0004-637X/762/1/26}, \href
  {https://ui.adsabs.harvard.edu/abs/2013ApJ...762...26Y} {762, 26}

\bibitem[\protect\citeauthoryear{{de los Reyes}, {Kirby}, {Seitenzahl}  \&
  {Shen}}{{de los Reyes} et~al.}{2020}]{DeLosReyes20}
{de los Reyes} M. A.~C.,  {Kirby} E.~N.,  {Seitenzahl} I.~R.,   {Shen} K.~J.,
  2020, \mn@doi [\apj] {10.3847/1538-4357/ab736f}, \href
  {https://ui.adsabs.harvard.edu/abs/2020ApJ...891...85D} {891, 85}

\makeatother
\end{thebibliography}





\bsp	
\label{lastpage}
\end{document}